\documentclass[twocolumn]{aastex63}

\usepackage{graphicx}
\usepackage{color, colortbl}
\usepackage{float}

\usepackage{lineno}

\shorttitle{Swift J1728.9-3613}
\shortauthors{Draghis et al.}

\begin{document}

\title{The Spin of a Newborn Black Hole: Swift J1728.9-3613}

\author[0000-0002-2218-2306]{Paul A. Draghis}
\email{pdraghis@umich.edu}
\affiliation{Department of Astronomy, University of Michigan, 1085 South University Avenue, Ann Arbor, MI 48109, USA}

\author[0000-0001-9641-6550]{Mayura Balakrishnan}
\affiliation{Department of Astronomy, University of Michigan, 1085 South University Avenue, Ann Arbor, MI 48109, USA}

\author[0000-0003-2869-7682]{Jon M. Miller}
\affiliation{Department of Astronomy, University of Michigan, 1085 South University Avenue, Ann Arbor, MI 48109, USA}

\author[0000-0002-8294-9281]{Edward Cackett}
\affiliation{Department of Physics \& Astronomy, Wayne State University, 666 W Hancock St, Detroit, MI 48201, USA}

\author[0000-0002-9378-4072]{Andrew C. Fabian}
\affiliation{Institute of Astronomy, University of Cambridge, Madingley Road, Cambridge CB3 OHA, UK}

\author[0000-0003-3124-2814]{James C. A. Miller-Jones}
\affiliation{International Centre for Radio Astronomy Research, Curtin University, GPO Box U1987, Perth, WA 6845, Australia}

\author[0000-0002-0940-6563]{Mason Ng}
\affiliation{MIT Kavli Institute for Astrophysics and Space Research, Massachusetts Institute of Technology, Cambridge, MA 02139, USA}

\author[0000-0002-7868-1622]{John C. Raymond}
\affiliation{Center for Astrophysics, Harvard \& Smithsonian, 60 Garden St., Cambridge, MA 02138, USA}

\author[0000-0003-1621-9392]{Mark Reynolds}
\affiliation{Department of Astronomy, University of Michigan, 1085 South University Avenue, Ann Arbor, MI 48109, USA}
\affiliation{Department of Astronomy, Ohio State University,  140 W 18th Avenue, Columbus, OH 43210, USA}

\author[0000-0002-0572-9613]{Abderahmen Zoghbi}
\affiliation{Department of Astronomy, University of Michigan, 1085 South University Avenue, Ann Arbor, MI 48109, USA}
\affiliation{Department of Astronomy, University of Maryland, College Park, MD 20742, USA}
\affiliation{CRESST II, NASA Goddard Space Flight Center, Greenbelt, MD 20771, USA}

\begin{abstract}
The origin and distribution of stellar-mass black hole spins are a rare window into the progenitor stars and supernova events that create them.  Swift J1728.9-3613 is an X-ray binary, likely associated with the supernova remnant G351.9-0.9 (Balakrishnan et al. 2023).  A NuSTAR X-ray spectrum of this source during its 2019 outburst reveals reflection from an accretion disk extending to the innermost stable circular orbit.  Modeling of the relativistic Doppler shifts and gravitational redshifts imprinted on the spectrum measures a dimensionless spin parameter of $a=0.86\pm0.02$ ($1\sigma$ confidence), a small inclination angle of the inner accretion disk $\theta<10$ degrees, and a sub-solar iron abundance in the disk $A_{\rm Fe}<0.84$.  This high spin value rules out a neutron star primary at the $5\;\sigma$ level of confidence. If the black hole is located in a still visible supernova remnant, it must be young. Therefore, we place a lower limit on the natal black hole spin of $a>0.82$, concluding that the black hole must have formed with a high spin. This demonstrates that black hole formation channels that leave a supernova remnant, and those that do not (e.g. Cyg X-1), can both lead to high natal spin with no requirement for subsequent accretion within the binary system.  Emerging disparities  between the population of high-spin black holes in X-ray binaries and the low-spin black holes that merge in gravitational wave events may therefore be explained in terms of different stellar conditions prior to collapse, rather than different environmental factors after formation.
\end{abstract}


\section{Introduction} \label{sec:intro}

Whether a neutron star (NS) or a black hole (BH) is produced in a stellar core-collapse event likely depends on factors such as the progenitor star’s mass, metallicity, and rotation.  Each of these factors is difficult to determine after the event. Additionally, binary interactions such as stripping of outer stellar layers, tidal spin-up, or the presence of common envelope phases likely also influences the nature of the newly formed compact object.  However, even more observationally elusive parameters are likely to be important, and may also help to determine the character of the compact object that is produced, and whether or not a supernova remnant (SNR) is also left behind.  These factors include the stellar angular momentum profile, the degree of convection and the magnetic field configuration within the progenitor, the rate of neutrino production in the collapse, the degree of post-bounce outward pressure if a proto-neutron star (PNS) is formed, the strength of disk winds originating from the accretion disk around a PNS, and the impact of jet-induced bubbles on initially unaffected outer layers of the star (\citealt{1999ApJ...524..262M}). 

The presence or absence of a SNR is a potential signature of distinct black hole formation channels.  If black holes form through “direct collapse” - without first forming a PNS - there may not be enough outward pressure to generate a SNR.  The black hole in the archetype X-ray binary Cygnus X-1 is consistent with this scenario (see, e.g., \citealt{2017IAUS..324..303M, 2003Sci...300.1119M,  2020MNRAS.491.2715B, 2015MNRAS.450.3289G, 2015MNRAS.453.2885R}).   
Alternatively, some theoretical treatments suggest that collapse events that form a PNS and SNR can then form a black hole if the progenitor mass and fallback mass are both high (\citealt{2001ApJ...550..410M}).  These scenarios may be tied to hypernova events and gamma-ray bursts, potentially leading to black holes with high spin (\citealt{2011LNP...812..153T}).  W49B may be an example of such a SNR, though there is not yet evidence of a black hole within it (\citealt{2013ApJ...764...50L}).  

Black hole spin may be an even more incisive probe of formation channels and the elusive inner properties of the progenitor star (the dimensionless spin parameter is given by $a=cJ/GM^2$, where $-0.998 \leq a \leq 0.998$, \citealt{1974ApJ...191..507T}). In order to form a rapidly rotating black hole, the compact object must either inherit angular momentum from the collapse of a rapidly rotating progenitor or accrete material with high angular momentum originating from the progenitor star. This accretion, whether immediate or with a small delay, is likely consistent with nearly solid-body rotation within the progenitor. This configuration may be natural in binary systems, where tidal interactions with the companion star must influence the rotation rate and structure of the progenitor (\citealt{2022arXiv220108407F}).  There may even be important secondary effects: rotation within the progenitor star serves to increase the maximum supported mass of the PNS before collapse to a BH, influencing the neutrino flux before collapse (\citealt{2021arXiv211209707R}) and driving up the natal spin of the black hole.

The distribution of black hole spins in X-ray binaries (XB) appears to be strongly skewed to high values (\citealt{2021arXiv211102935F, 2022arXiv221002479D}).  However, while the BHs in high-mass X-ray binaries (HMXB) must have been formed with the high spins that we see due to their short lifetimes (\citealt{1976IAUS...73...35V, 1995ApJS..100..217I}), it is possible that in the case of low-mass X-ray binaries (LMXB) the values measured today do not reflect natal spins, but instead prolonged episodes of super-Eddington accretion (\citealt{2015ApJ...800...17F}). Resolving this disparity is not only central to understanding black hole formation channels, but central to understanding the origins of the merging low-spin black holes that are inferred in gravitational wave events (\citealt{2021arXiv211103634T}).

The two preferred methods for measuring black hole spin using X-ray spectral observations are continuum fitting (see e.g., \citealt{2009ApJ...701.1076G, 2014ApJ...784L..18M, 2020MNRAS.499.5891S}) and relativistic reflection  (see e.g., \citealt{2006ApJ...652.1028B, 2007ARA&A..45..441M, 2014SSRv..183..277R, 2021SSRv..217...65B}).  The relativistic reflection method is independent of black hole mass, distance, and accretion rate, so it is suited to measuring spin in systems that do not have prior constraints on these parameters. The inclination of the inner accretion disk is treated as a free parameter and could, in principle, be different from the orbital inclination due to disk tearing and the Bardeen-Petterson effect (\citealt{1975ApJ...195L..65B, 2015MNRAS.448.1526N, 2021MNRAS.507..983L}).

The main feature of reflection is an Fe K emission line  that is produced when one of the two K-shell electrons of an Fe atom is ejected by an ionizing X-ray photon (produced in a compact, central corona).  Relativistic Doppler shifts and gravitational red-shifts alter the shape of the Fe K line by ``blurring" it, with the effects becoming stronger with proximity to the black hole.   Because the innermost stable circular orbit (ISCO) is set by the spin of the black hole (\citealt{1972ApJ...178..347B, 1973blho.conf..343N}), measuring the shape of the Fe K line offers a direct way to measure the black hole spin.  While soft X-rays are preferentially absorbed by the disk, the hard part of the incident flux from the corona is preferentially Compton scattered. This gives the reflection spectrum above $\sim20~\rm keV$ a broad shape, commonly known as the ``Compton hump". Due to its wide pass band spanning between $3-79~\rm keV$, its increased sensitivity and its ability to observe bright sources without pile-up effects, NuSTAR (\citealt{2013ApJ...770..103H}) is an ideal instrument for measuring the effects of relativistic disk reflection and black hole spin. 


The geometry of the hard X-ray corona has an important impact on the expected spectrum.  Well-developed models of X-ray reflection divide into two flavors: those that assume a ``lamp-post" geometry, and those that parameterize the geometry in terms of radial emissivity indices that can be determined by the data.  A key assumption underlying all black hole spin measurement methods, including relativistic reflection modeling, is that of an optically thick, geometrically thin accretion disk (\citealt{1973A&A....24..337S}) that extends all the way to the ISCO of the black hole, and that any matter closer to the black hole is on plunging orbits and cannot contribute significantly to the total emission. This assumption is supported by the results of numerical simulations that find a sharp inner disk boundary exists for Eddington fractions between $10^{-3}\lesssim L_{disk}/L_{Edd}\lesssim 0.3$ (see e.g., \citealt{2008ApJ...675.1048R, 2008ApJ...676..549S, 2016ApJ...819...48S}). In addition, this choice is motivated by numerous previous observational results confirming a constant inner disk radius, consistent with the ISCO (see e.g., \citealt{2010ApJ...718L.117S, 2013MNRAS.431.3510S, 2015ApJ...813...84G}).

Recently, Balakrishnan et al. 2023 identified Swift J1728.9-3613 as a black hole X-ray binary probably associated with the supernova remnant G351.9-0.9. While the mass of the black hole is unconstrained, the estimated distance to Swift J1728.9-3613 is $8.4\pm0.8\;\rm kpc$.  As the supernova remnant is observed to be in the Sedov phase, its age is likely $\lesssim 30,000$ years. If the explosion that produced the supernova remnant G351.9-0.9 also formed the black hole in Swift J1728.9-3631, then the black hole in this system is young, and a measurement of its spin provides a rare insight into natal black hole spin, the mechanisms governing supernovae explosions, the early stages of black hole formation, and also the physics of X-ray binaries. Therefore, we use relativistic reflection modeling to measure the spin of the BH in Swift J1728.9-3613, based on a NuSTAR observation taken during its 2019 outburst. In Section \ref{sec:obs} we present the spectral extraction process, in Section \ref{sec:analysis} we present our analysis and results, we interpret our findings, and discuss the possible implications of our results in Section \ref{sec:disc}.

\section{Observations and Data Reduction} \label{sec:obs}
\begin{figure}[ht!]
    \centering
    \includegraphics[width= 0.45\textwidth]{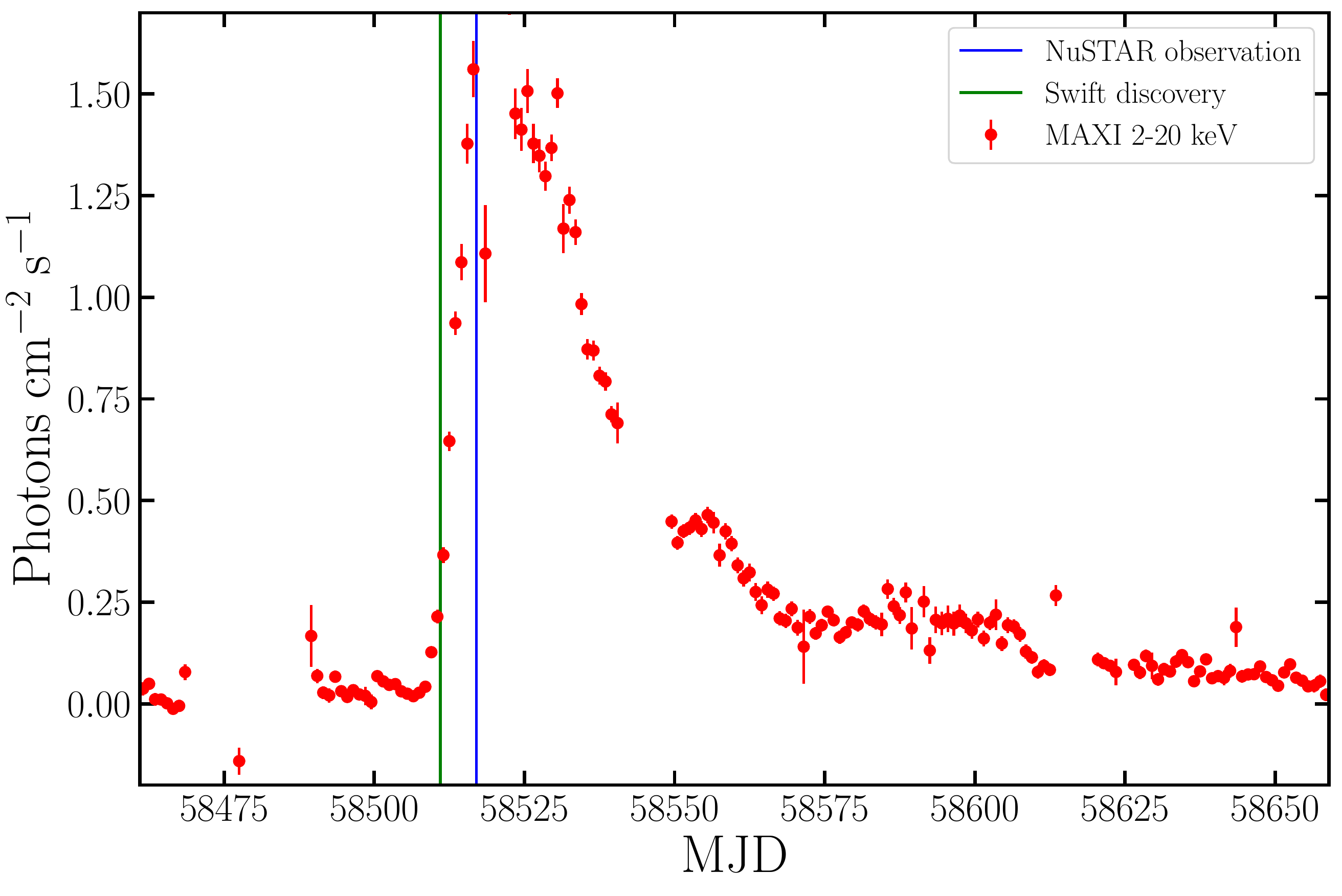}
    \caption{MAXI light curve of the 2019 outburst of Swift J1728.9-3613 in the 2-20~keV band, represented by the red points. The vertical green line represents the date of the Swift detection, and the vertical blue line shows the time of the NuSTAR observation analyzed by this work. NuSTAR observed the source before the peak of the outburst.}
    \label{fig:maxi}
\end{figure}

NuSTAR observed Swift J1728.9-3613 on February 3rd 2019 starting at 15:01:09 UT under ObsID 90501303002 for a net exposure of 16.7 ks, obtaining an average of $\sim$600 counts/second when combining the two NuSTAR FPM sensors. Figure \ref{fig:maxi} shows the MAXI (\citealt{2009PASJ...61..999M}) light curve of Swift J1728.9-3613 throughout its 2019 outburst. The red points represent 24h monitoring time bins in the 2-20~keV band. The vertical green line represents the initial Swift BAT detection (\citealt{2019ATel12436....1B, 2019ATel12445....1K}), and the blue line represents the NuSTAR observation, which occurred 9 days after the initial discovery of Swift J1728.9-3613, but still before the peak of the outburst.

\begin{figure}[ht!]
    \centering
    \includegraphics[width= 0.45\textwidth]{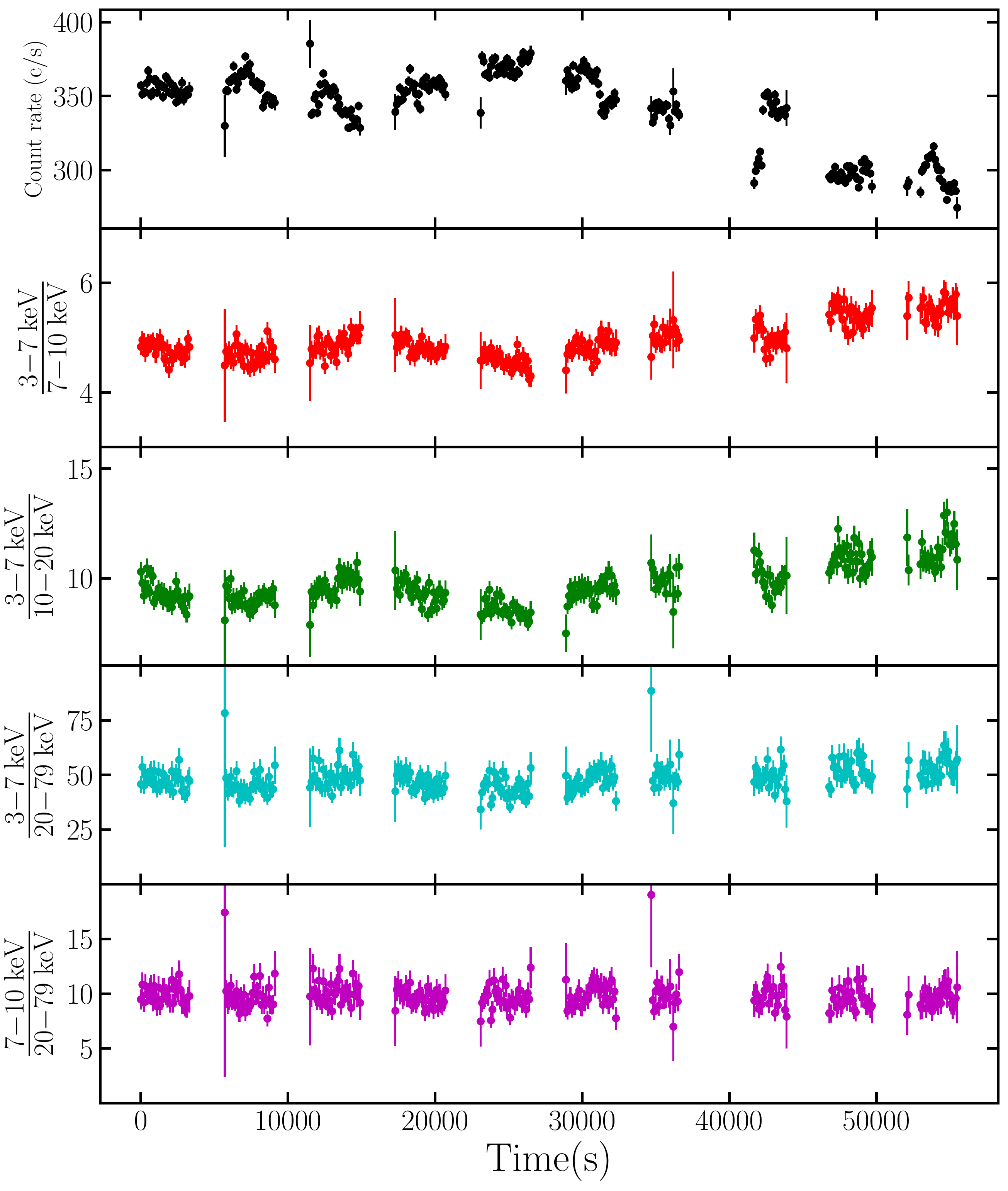}
    \caption{Light curve of the NuSTAR observation. The top panel shows the count rate observed by the FPMA NuSTAR detector across the entire 3--79~keV band pass. The second, third, and fourth panels show the evolution throughout the observation of the ratios of the count rate in the 3-7~keV band to that in the 7--10~keV, 10--20~keV, and 20--79~keV bands respectively. The last panel shows the evolution of the ratio of the count rate in the 7--10~keV band to that in the 20--79~keV band. Each point represents a time bin of 100 seconds.}
    \label{fig:light_curve}
\end{figure}

The top panel in Figure \ref{fig:light_curve} shows the light curve of the NuSTAR observation, as seen by FPMA, with each point representing a time bin of 100 seconds. Toward the end of the exposure, the count rate drops by $\sim 15\%$ for both detectors. Still, as the observation took place before the peak of the outburst and since the flux of the source would continue to increase over the following days, we considered this flux reduction to simply be attributed to random fluctuations around the trend. The second, third, and fourth panels in Figure \ref{fig:light_curve} show the evolution throughout the observation of the ratios of the count rate in the 3-7~keV band to that in the 7-10~keV, 10-20~keV, and 20-79~keV bands respectively. The last panel shows the evolution of the ratio of the count rate in the 7-10~keV band to that in the 20-79~keV band. These panels illustrate that the hardness of the source does not vary significantly throughout the duration of the NuSTAR exposure. Therefore, we extracted time-averaged spectra using the routines in HEASOFT v6.28 through the NuSTARDAS pipeline v2.0.0 and CALDB v20210524. The spectra were extracted from circular source regions with 120'' radii centered at the position of the source on the two FPM detectors. Background events were extracted from regions of the same size as the source regions. The spectra were grouped through the optimal binning scheme described by \cite{2016A&A...587A.151K}, using the ``ftgrouppha" ftool. 

In the case of this observation, the spectra from the two NuSTAR FPM detectors show a difference at low energies. \cite{2020arXiv200500569M} describes a soft excess in the FPMA NuSTAR sensor due to a tear in its Multi Layer Insulation (MLI) thermal blanket. In this case, we notice a decrease in flux in the FPMA spectrum. Since this effect is accounted for by CALDB in the spectrum extraction process, we tested whether correction for the low energy difference was overestimated by re-extracting the spectra using the CALDB version previous to the introduction of the MLI correction. The spectra showed the same low energy difference, indicating that the difference was not induced by a CALDB over-correction. We also tested whether the size and sub-pixel position of the source and background extraction regions had an influence over the low energy difference between the spectra from the two FPM detectors. We tested an array or region sizes (60'', 80'', 100'', 120'', and 200'' radii) to probe for the source of the difference, but the spectra were virtually identical for the five different cases. Additionally, we tested slightly adjusting the position of the source regions or shifting the position of the background regions and using an annular region centered at the position of the source instead of a circular extraction region. None of these experiments reduced the difference between the spectra below 4~keV. If we ignore the data below 5 keV, the reflection component of the best fit remains unchanged. Therefore, throughout our analysis we continued using the entire 3--79~keV NuSTAR pass band.

\section{Analysis and Results} \label{sec:analysis}

\begin{figure}[ht!]
    \centering
    \includegraphics[width= 0.45\textwidth]{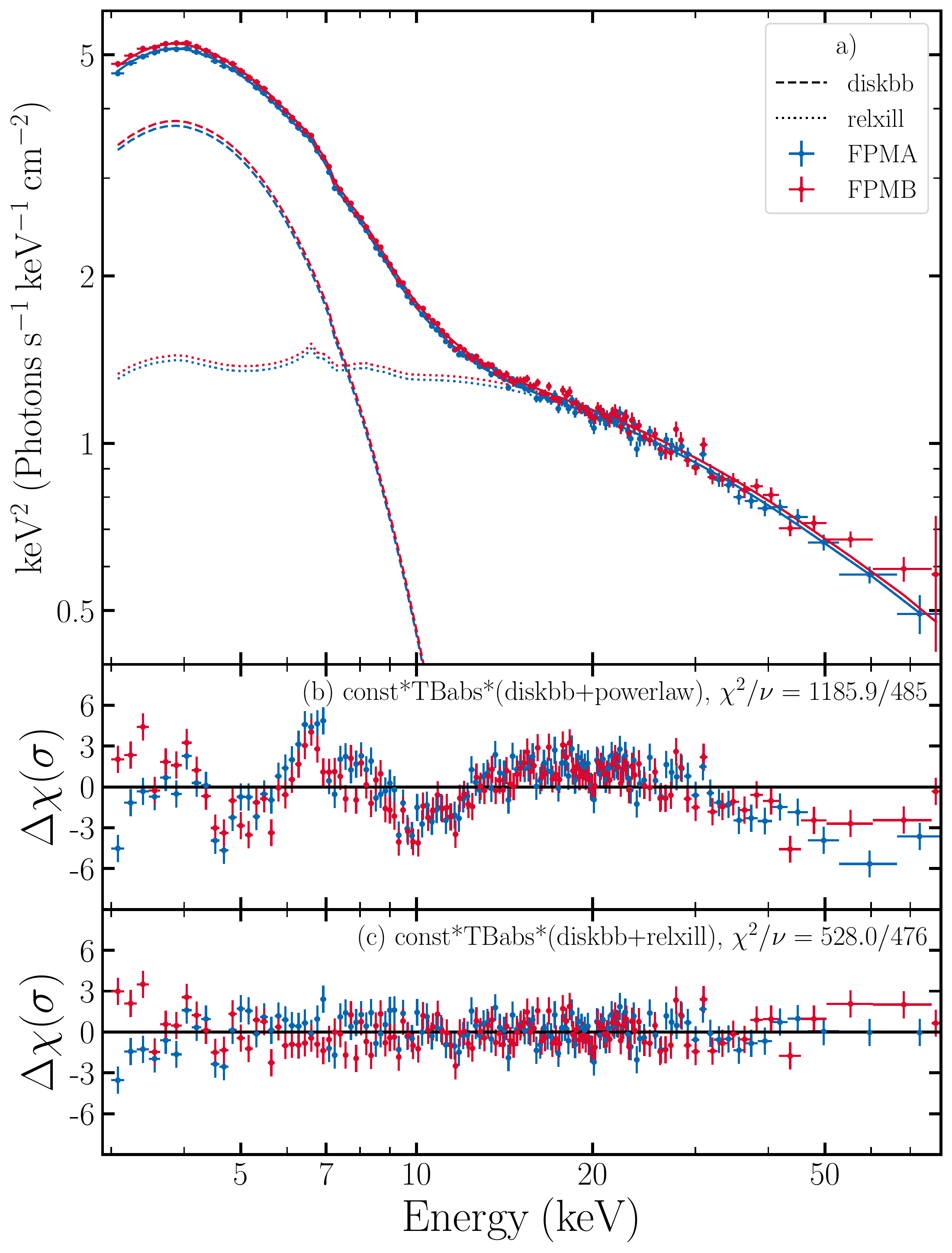}
    \caption{Panel (a) shows the unfolded spectrum of Swift J1728.9-3613. The blue points represent the spectrum from the NuSTAR FPMA detector, while the red points show the spectrum obtained from the FPMB sensor. The solid blue and red lines show the total model to the FPMA and FPMB spectra, while the dashed and dotted show the contribution to the model by the \texttt{diskbb} and \texttt{relxill} components, respectively.  Panel (b) shows the residuals in terms of $\sigma$ for the \texttt{constant*TBabs*(diskbb+powerlaw)} model. The residuals show clear indication of relativistic reflection. Panel (c) shows the residuals of the best-fit model, \texttt{constant*TBabs*(diskbb+relxill)}.}
    \label{fig:delchi}
\end{figure}

Spectral fitting was run using XSPEC v12.11.1d (\citealt{1996ASPC..101...17A}) and the quality of the fits was quantified using the $\chi^2$ statistic. Initial fits to the spectra were run with the combination of a \texttt{diskbb} component (\citealt{1984PASJ...36..741M}) describing the radiation from an optically thick, geometrically thin accretion disk and a power law component describing the emission from a compact corona. Those fits return a $\chi^2/\nu=1185.9/485=2.45$, where $\nu$ is the number of degrees of freedom. This model also includes the multiplicative component \texttt{TBabs} (\citealt{2006HEAD....9.1360W}) to account for the interstellar absorption using abundances computed by \cite{2000ApJ...542..914W} and photoionization cross-sections computed by \cite{1996ApJ...465..487V}. In addition, the model contains a multiplicative constant between the spectra obtained from the two NuSTAR FPM sensors, and all other parameters of the model are linked between two spectra. This constant takes a value $\sim1.02$, in accordance with known instrument calibration uncertainties. The residuals of this fit are presented in panel b) of Figure \ref{fig:delchi}, showing strong signs of relativistic disk reflection including the broadened Fe K line around 6.4 keV and the Compton hump above 20 keV.

\subsection{Best Model}\label{sec:best_model}

In order to account for the reflection features, we replaced the \texttt{powerlaw} component with variations of the \texttt{relxill} v1.4.3 model (\citealt{2014MNRAS.444L.100D, 2014ApJ...782...76G}). The \texttt{relxill} family of models has become a standard in relativistic reflection modeling by combining reflection with relativistic broadening effects. We explain the specific details of the \texttt{relxill} family of models in Appendix \ref{sec:relxill}, together with our general assumptions regarding model parameters. The best model in terms of $\chi^2$ was obtained by replacing the \texttt{power-law} component of the model shown in panel b) in Figure \ref{fig:delchi} with the basic version of \texttt{relxill}, which makes no assumptions about the coronal geometry, has the disk density fixed at $\log(n)=15$, and models the illuminating flux as a high energy cutoff power law. The full model thus becomes \texttt{const*TBabs*(diskbb+relxill)}. The quality of the fit is significantly improved over the previous model, with $\chi^2/\nu=528/476=1.11$. The residuals of this model are shown in panel c) of Figure \ref{fig:delchi}. Panel a) of Figure \ref{fig:delchi} shows the unfolded spectrum of Swift J1728.9-3613 and the model components. Blue points represent the spectrum obtained from the FPMA NuSTAR sensor, while red points show the FPMB spectrum. The complete model is represented by the solid line, while the dashed and dotted lines show the contributions of the \texttt{diskbb} and \texttt{relxill} components respectively. We note that most of the contribution to $\chi^2$ at this point comes from the difference between the spectra from the two NuSTAR sensors at energies below 4~keV, discussed in Section \ref{sec:obs}. This difference is best noticeable in panel c) of Figure \ref{fig:delchi}.

When testing the ``lamp post" geometry by replacing the \texttt{relxill} component with \texttt{relxilllp}, the fit returns $\chi^2/\nu=615.4/478=1.29$, worse by $\Delta\chi^2=87.4$ for $\Delta \nu=2$. The spin of the compact object, the inclination of the inner disk, and the height of the corona are unconstrained by the fit. Replacing the \texttt{relxill} component with \texttt{relxillCp} which replaces the high energy cutoff power law incident flux with that of a Comptonization continuum decreases the quality of the fit by $\Delta \chi^2=25$ for no change in number of degrees of freedom. Lastly, replacing the \texttt{relxill} component with \texttt{relxillD} worsens the fit by $\Delta \chi^2\approx 3$, with $\chi^2$ becoming progressively worse by an additional $\Delta \chi^2\approx 2$ as the disk density is changed between the hard limits of $\log(n)=15$ and $\log(n)=19$. Despite \texttt{relxill} having a fixed value of $\log(n)=15$, replacing it with \texttt{relxillD} and fixing $\log(n)=15$ worsens the fit because \texttt{relxillD} fixes $E_{\rm cut}=300\;\rm keV$, while \texttt{relxill} allows $E_{\rm cut}$ to vary. With the quality of the fit slightly worsening with increasing disk density, one would assume that an even smaller disk density would produce a better fit. In practice, the small differences in the quality of fit likely indicate that when using the \texttt{relxillD} component to fit the spectra, the quality of the fit is not strongly sensitive to the disk density. An alternative relativistic reflection model which can probe higher disk densities up to $\log(n)=22$ is \texttt{reflionx\_HD} (see e.g., \citealt{2020MNRAS.498.3888J, 2021ApJ...909..146C}). We fit the NuSTAR spectra with \texttt{constant*TBabs* (diskbb+relconv*reflionx\_HD)}, but the quality of the fit is significantly worse than out baseline fit: $\chi^2/\nu=682.57/475=1.44$, and the disk density takes a value of $n\sim 10^{17} \; \rm cm^{-3}$ but poorly constrained, and the BH spin, viewing inclination, disk ionization and Fe abundance remain relatively unchanged, well consistent with the values measured using \texttt{relxill}. When fixing the disk density to $n\sim 10^{21} \; \rm cm^{-3}$, the fit becomes slightly worse, suggesting that the fits are mostly insensitive to the assumed disk density, as most of the effects of this parameter would produce effects mainly noticeable at energies below the NuSTAR band pass. Nevertheless, we note that as we change the assumed disk density, the spin and inclination measurements are not strongly impacted. Therefore, for the rest of the analysis, we continued using the model containing the basic version of \texttt{relxill}, illustrated in panel a) of Figure \ref{fig:delchi}. Lastly, we note that the choice of model used to account for the emission from the accretion disk does not influence the reflection features and the spin and inclination measurement. We discuss the effects of different disk models in Appendix \ref{sec:corner}.

\begin{figure}[ht!]
    \centering
    \includegraphics[width=0.45\textwidth]{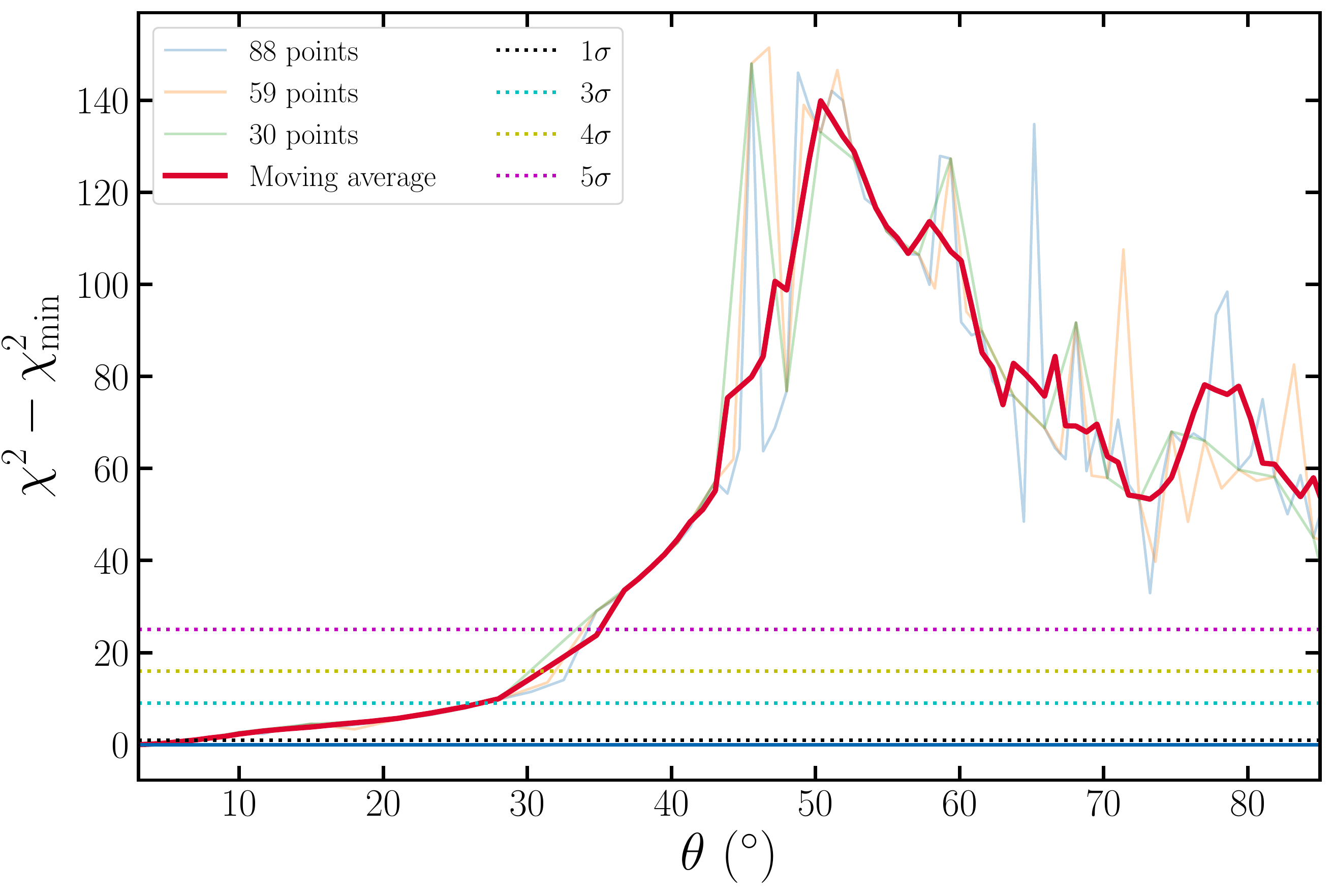}
    \caption{Contour showing the results of the \texttt{steppar} function for the inclination parameter. The y axis was shifted to show the increase in $\chi^2$ when compared with the best-fit model. The thin green, orange, and blue lines show the output when the function evaluates the statistic over at points that are oversampling the \texttt{relxill} tables by a factor of 1, 2, and 3, located over the (3,85) parameter space for the inclination. The solid red line represents a moving average of the 88-point run over a 5-point averaging window. The horizontal solid blue line represents the $\chi^2=\chi^2_{\rm min}$ contour, while the dotted black, cyan, yellow, and magenta lines represent $1\sigma$, $3\sigma$, $4\sigma$, and $5\sigma$ significance contours.}
    \label{fig:incl_steppar}
\end{figure}

In our best-fit model, the measured inclination of the inner accretion disk is low, taking a value around $\theta\sim4^\circ$, but the Xspec fitting algorithm \texttt{leven} is unable to place constraints on its uncertainty. We ran the command \texttt{steppar} on the inclination parameter between the limits allowed by \texttt{relxill} - 3 and 85 degrees - evaluated at 30, 59, and 88 inclination values, corresponding to 1, 2, and 3 times the sampling of the \texttt{relxill} tables of the inclination parameter space. This function performs a fit at each step and computes the value of $\chi^2$. The output of this experiment is shown in Figure \ref{fig:incl_steppar}. The blue line represents the minimal value of $\chi^2=528$ obtained for $\rm \theta=3.6^\circ$. The outputs of the \texttt{steppar} runs are shown by the thin green, orange, and green lines respectively. The solid red line shows the moving average of the \texttt{steppar} run evaluating the statistic at 88 points, binning the values in a 5 point width window. The horizontal dotted black, cyan, yellow, and magenta lines represent the difference in $\chi^2$ corresponding to $1\sigma$, $3\sigma$, $4\sigma$, and $5\sigma$. This constrains the inclination angle of the system to low values, with $\rm \theta\lesssim35^\circ$ at a $5\sigma$ confidence level. We further explore the robustness of our low-inclination measurement and discuss its implications in Appendix \ref{sec:incl}.

\subsection{Markov Chain Monte Carlo (MCMC) Analysis}\label{sec:MCMC}
In order to probe the shape of the parameter space and determine the uncertainties of our measurements, we used the best fit parameter combination to generate a proposal distribution for a Markov Chain Monte Carlo (MCMC) algorithm. We ran the algorithm with 200 walkers for a total of $10^7$ steps, using the XSPEC EMCEE implementation written by A. Zoghbi\footnote{ \url{https://zoghbi-a.github.io/xspec\_emcee/} }, with the first $2\times10^6$ steps being considered a ``burn in'' phase and disregarded in future analysis. For further explanation on the choice of chain parameters, see Section 3.2 in \cite{2021ApJ...920...88D}. We generated corner plots from the samples using the \emph{corner} Python module (\citealt{corner}). Appendix \ref{sec:corner} contains the complete corner plot, including all the free parameters in the analysis and discusses a few possible correlations on parameters and their effect on the spin measurement.

For parameters such as $q_1$, $R_{\rm br}$, or $\rm \theta$ the median of the posterior distribution is significantly skewed from the mode and not entirely representative of the shape of the posterior distribution. Therefore, we chose to report the mode of the posterior distribution for each parameter in the MCMC analysis and the $1 \sigma$ credible region of the distribution, representing the upper and lower limits of the minimum length interval containing $68.3\%$ of the posterior distribution samples. These values are reported in Table \ref{table:results}. For parameters indicated by $\star$, one of the boundaries of the $1 \sigma$ credible region coincides with the upper or lower limit that the parameter can take in the model.

The measured disk temperature $kT_{\rm in}=1.28\pm0.01\;\rm keV$ is high, similar to those previously been measured in MAXI J1535-571 (\citealt{2018ApJ...860L..28M}) or XTE J1550-564 (\citealt{2000ApJ...544..993S}).  Often, similar fits to black hole spectra with \texttt{relxill} variants require enhanced Fe abundances, likely due to a degeneracy with the disk density (\citealt{2018ApJ...855....3T}). In this case, the sub-solar iron abundance $A_{\rm Fe} < 0.84$ (in units of the Solar abundance - \citealt{1996ASPC...99..117G}) in the accretion disk is indicative of a low metallicity companion. The power law index $\Gamma=2.21\pm0.03$ and ionization $Log(\xi)=3.6\pm0.1$ are similar to those measured in other systems (see e.g., \citealt{2020ApJ...900...78D, 2021ApJ...920...88D}). Given the definition of the ionization parameter  presented in section \ref{sec:relxill} and based on the size of the inner disk radius and disk density, one can compute the expected ionizing luminosity. In this case, assuming a $10\;M_\odot$ BH, and $r\sim10\; r_{\rm g}$, then $L\sim10^{33}\;\rm erg/s$, while the Eddington luminosity would be on the order of $L_{Edd}\sim10^{39}\;\rm erg/s$. The most straightforward choice for solving this discrepancy of 5-6 orders of magnitude between the computed ionized flux and the expected flux from a BH of this size would be adopting a model that uses a higher disk density. Still, as mentioned in section \ref{sec:best_model}, models that use the \texttt{relxillD} flavor which allows for higher disk densities produce slightly worse fits and similar values of the ionization parameter. It is important to note that \texttt{relxillD} only allows for disk densities up to $n=10^{19}\;\rm cm^{-3}$, while works such as \cite{2021ApJ...909..146C} show that the expected densities are $n\geq10^{20}\;\rm cm^{-3}$ for similar accretion disks. Additionally, \cite{2021ApJ...909..146C} point out that the reported value for the ionization parameter is an average across the disk, and if the disk is not truncated and the corona is compact (both expected in this case), then the flux in the inner regions of the disk can be much higher than a simple estimate using the definition of the ionization parameter.

\begin{figure}[ht!]
    \centering
    \includegraphics[width= 0.45 \textwidth]{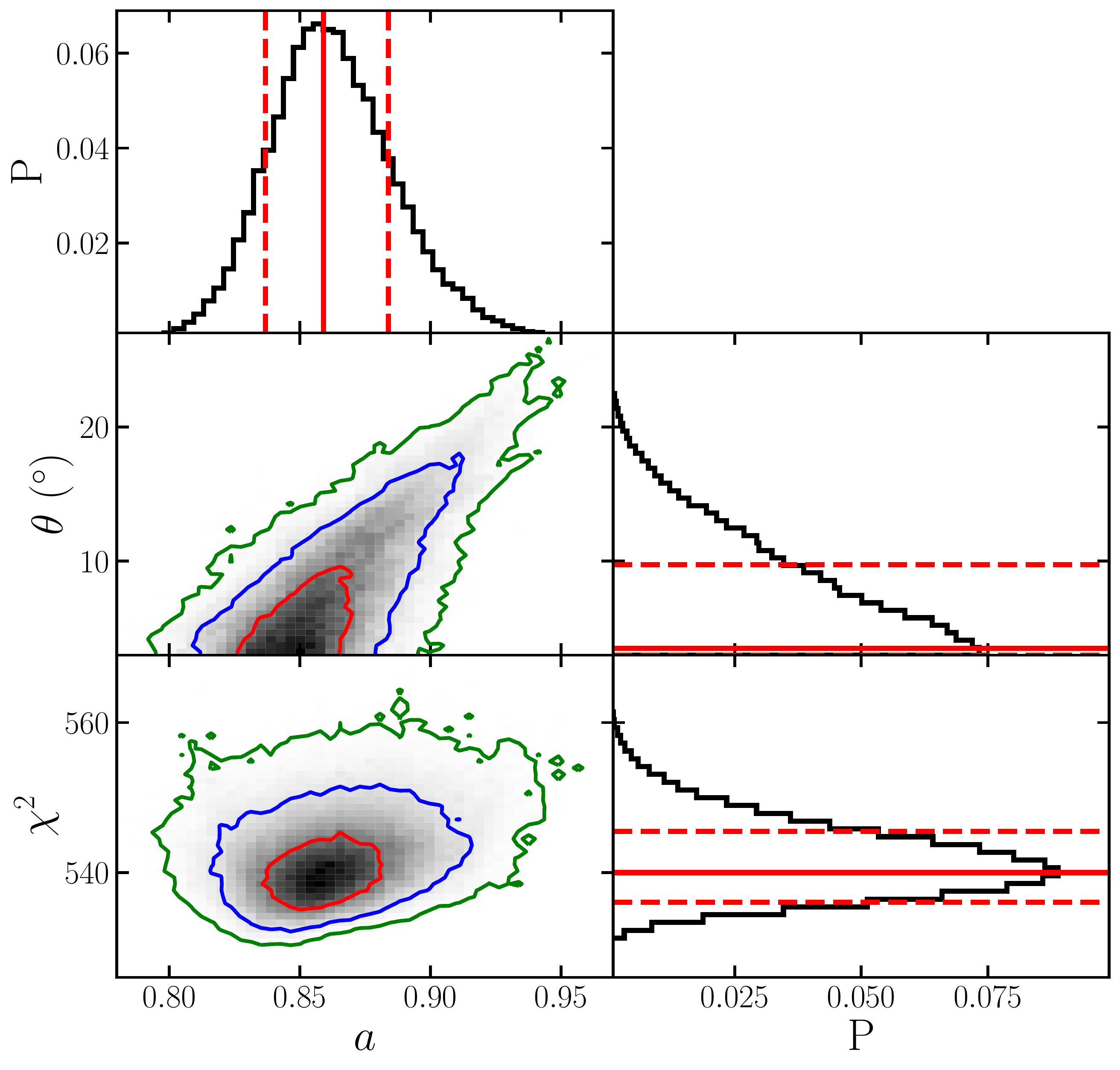}
    \caption{2D histograms of the $a$-$\theta$ (center left panel) and of the $a$-$\chi^2$ (bottom left panel) parameter space based on the posterior samples in the MCMC analysis. The red, blue, and green contours in these panels represent the $1\sigma$, $2\sigma$, and $3\sigma$ confidence intervals respectively. The top left, middle right, and bottom right panels show the 1D histograms of the posterior distribution in the MCMC analysis for spin, inclination, and $\chi^2$. The solid red lines represent the mode of the distributions and the dashed red lines represent the $\pm1\sigma$ credible regions.}
    \label{fig:spin_incl_chi}
\end{figure}

The cutoff energy of the incident power law $E_{cut} > 650~\rm keV$ is high and poorly constrained, with values close to the upper limit of $1000~\rm keV$ being strongly preferred in the MCMC analysis. The reflection fraction $\rm R=3.2\pm0.6$ is also high, but for the measured spin it is within the maximum possible value predicted by \cite{2014MNRAS.444L.100D}. If we try to fix the reflection fraction to $\rm R=1$, the fit worsens by $\Delta \chi^2=5$, the Fe abundance stays low but poorly constrained $A_{Fe}\sim1$ and the spin prediction remains unchanged. The inner emissivity profile is steep, with $q_1>8.7$ up to $R_{\rm br}=8^{+3}_{-1}\;r_{\rm g}$, followed by a flattening to $q_2=1.6^{+0.3}_{-0.8}$. This is in accordance with the theoretical predictions of \cite{2012MNRAS.424.1284W}, which suggest a similar behavior over the inner regions of the accretion disk. While the theoretical expectation is that the emissivity should take a constant index of $q_{\rm out}=3$ over the outer regions of the accretion disk, current models do not include the possibility of an emissivity profile with more than two steps. Still, based on this emissivity profile, the outer regions of the accretion disk are not expected to contribute significantly to the observed X-ray flux. We note that when fixing $q_1=q_2=3$, the fit becomes worse by $\Delta \chi^2\sim120$ and the spin prediction remains high but much worse constrained. 

\begin{deluxetable}{c|cc}
\tablecaption{Results of the MCMC analysis}
\label{table:results}
\tablewidth{\textwidth} 
\tabletypesize{\scriptsize}
\tablehead{
\colhead{Component} & \colhead{Parameter} & \colhead{Value}
}
\startdata\\
$\texttt{constant}$ & $\rm Const$ & $1.020^{+0.001}_{-0.001}$ \\\hline
$\texttt{TBabs}$ & $N_{\rm H} \: [\rm \times 10^{22} \: cm^{-2}]$ & $3.7^{+0.2}_{-0.1}$ \\\hline
$\texttt{diskbb}$ & $kT_{\rm in} \: [\rm keV]$ & $1.28^{+0.01}_{-0.01}$ \\
 & $\rm norm_{\rm d}$ & $260^{+8}_{-17}$ \\\hline
$\texttt{relxill}$ & $q_{1}$ & $\star9.9^{+0.1}_{-1.2}$ \\
 & $q_{2}$ & $1.6^{+0.3}_{-0.8}$ \\
 & $R_{\rm br}\:[r_g]$ & $8^{+3}_{-1}$ \\
 & $a$ & $0.86^{+0.02}_{-0.02}$ \\
 & $\rm \theta \:[^\circ]$ & $\star3.5^{+6.2}_{-0.5}$ \\
 & $\Gamma$ & $2.21^{+0.03}_{-0.03}$ \\
 & $\log(\xi)\;[\rm erg\;cm\;s^{-1}]$ & $3.6^{+0.1}_{-0.1}$ \\
 & $A_{\rm Fe}\;[A_\odot]$ & $\star0.53^{+0.31}_{-0.03}$ \\
 & $E_{\rm cut}\: [\rm keV]$ & $\star960^{+40}_{-310}$ \\
 & $\rm R$ & $3.2^{+0.6}_{-0.6}$ \\
 & $\rm norm_{\rm rel} \: [\rm \times 10^{-3}]$ & $10^{+2}_{-1}$ \\\hline
 & $\chi^2 / \nu$ & $534^{+6}_{-4} \; (528)/476 \;=\;1.11 $\\\hline
\enddata
\tablecomments{In this table, we report the mode of the posterior distributions in the MCMC analysis, along with the $1\sigma$ credible region. Parameters marked with $\star$ indicate that one of the limits of the $1\sigma$ credible region overlaps with the hard limit of the parameter in the model. For $\chi^2$, the number in parentheses indicates the best-fit $\chi^2$ value.}
\end{deluxetable}

The measured inclination is low, $\rm \theta \lesssim 10^\circ$, with lower values being strongly favored (see the center-right panel of Figure \ref{fig:spin_incl_chi}). The spin is high and well constrained, $a=0.86\pm0.02$. Figure \ref{fig:spin_incl_chi} shows the distribution of the posterior samples in the inclination-spin and $\chi^2$-spin space in the center and lower left panels respectively, with the red, blue, and green contours representing $1\sigma$, $2\sigma$, and $3\sigma$ regions respectively. The top left, center right, and bottom right panels show the 1D posterior distributions for spin, inclination, and $\chi^2$ respectively, in the posterior MCMC distributions. The solid red lines in these panels represent the mode of the posterior distributions and the $68.3\%$ credible regions, reported in Table \ref{table:results}. The posterior samples of the MCMC analysis show a positive correlation between the spin of the compact object $a$ and the inclination of the inner accretion disk $\rm \theta$ (see the center-left panel of Figure \ref{fig:spin_incl_chi}). This can be explained through the fact that both gravitational redshifts and relativistic Doppler shifts produce line asymmetry. Still, in our posterior sample the inclination only takes values as high as $30^\circ$ and, as we have shown in section \ref{sec:best_model} and explored further in Appendix \ref{sec:incl}, the preferred inclination is low and therefore this partial degeneracy between the two parameters does not significantly influence the spin measurement. As shown in Appendix \ref{sec:incl}, higher inclinations are highly disfavored, solidifying the confidence of the spin measurement.

\newpage
\subsection{Black Hole vs. Neutron Star}

\begin{figure}[ht!]
    \centering
    \includegraphics[width= 0.45\textwidth]{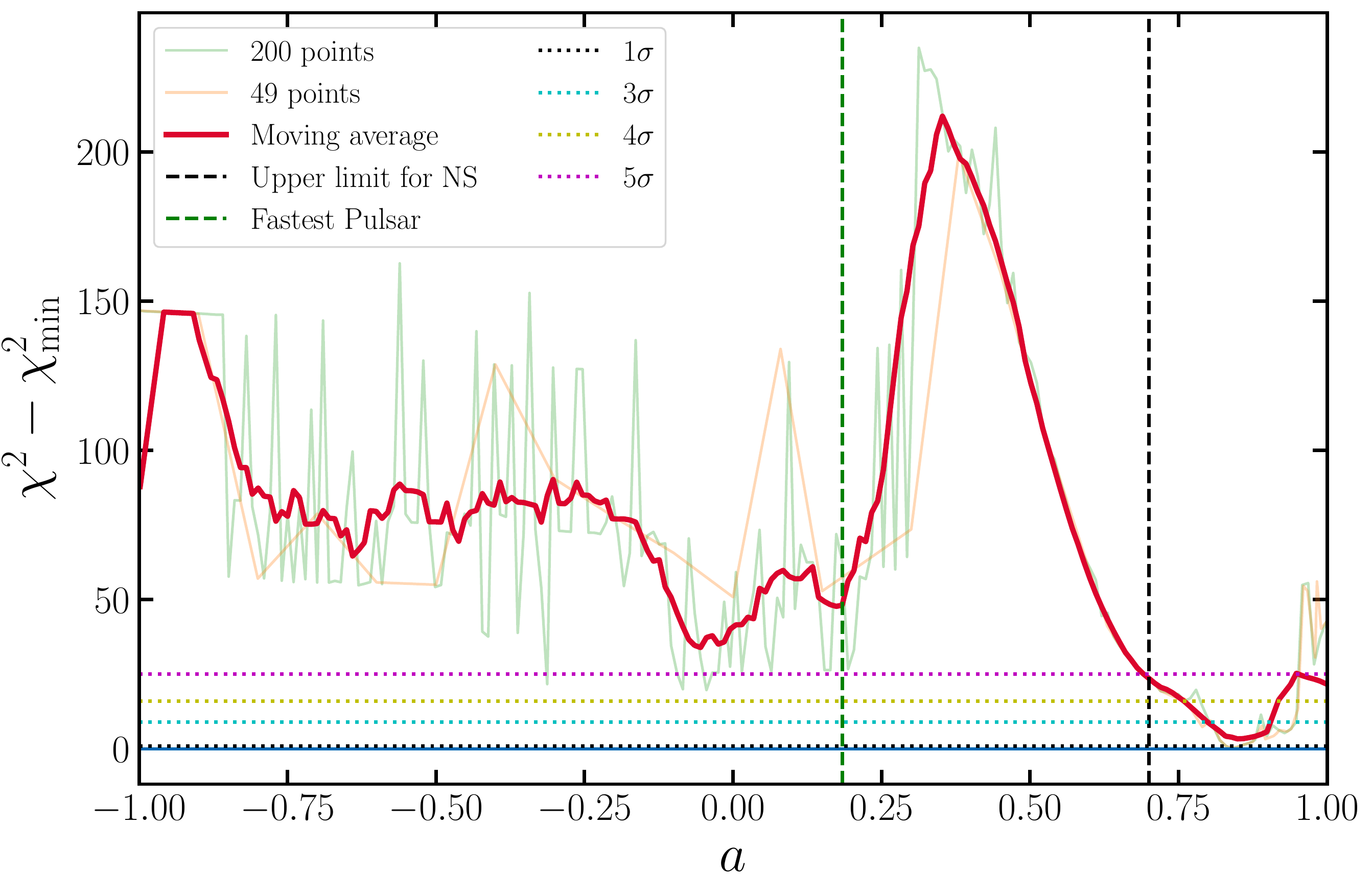}
    \caption{Contour showing the results of the \texttt{steppar} function for the Spin parameter. The y axis was shifted to show the increase in $\chi^2$ over the best-fit model. The thin green line shows the output when the function evaluates the statistic over 200 points equidistantly located over the (-0.998,0.998) parameter space for the spin. The thin orange line shows the output of the \texttt{steppar} run which evaluated the statistic at points that are oversampling the \texttt{relxill} tables by a factor of 2. The solid red line represents a moving average of the 200-point run over a 5-point averaging window. The horizontal solid blue line represents the $\chi^2=\chi^2_{\rm min}$ contour, while the dotted black, cyan, yellow, and magenta lines represent $1\sigma$, $3\sigma$, $4\sigma$, and $5\sigma$ significance contours. The green vertical dashed line shows the spin of the fastest known millisecond pulsar, while the black vertical dashed line shows the theoretical upper limit for the spin of a neutron star. The contour shows that the possibility of the compact object being a neutron star is excluded at more than $5\sigma$ confidence.}
    \label{fig:spin_steppar}
\end{figure}

Similarly to the inclination \texttt{steppar} experiment presented in Subsection \ref{sec:best_model}, we ran the \texttt{steppar} function on the spin parameter between the limits allowed by the model. The results are shown in Figure \ref{fig:spin_steppar}. 
The solid red line shows the moving average evaluated over a window of width equal to 5 points of a \texttt{steppar} run that was evaluating the $\chi^2$ at 200 points evenly spaced out between $a=-0.998$ and $a=0.998$.
The vertical dashed green line shows the spin value of the fastest millisecond pulsar, PSR J1748-2446ad ($\nu=716\;Hz$ - \citealt{2006Sci...311.1901H}), computed assuming a neutron star mass of $1.4M_\odot$ and a radius of $20~\rm km$ and assuming that the neutron star is a uniform, solid, rotating sphere. This combination of parameters gives $a=0.184$, with higher NS masses or lower radii reducing the value of the dimensionless spin. Using the more accurate moment of inertia of a neutron star approximated in equation 12 in \cite{2005ApJ...629..979L} and the same combination of parameters, we obtain an even lower spin for the fastest millisecond pulsar, $a=0.141$. The vertical dashed black line shows the maximum theoretical limit on the spin of a neutron star of $a=0.7$ (\citealt{2011ApJ...728...12L}). The "best fit" model predicts a high spin value for the central object, $a\sim0.85$, higher than any value possible for neutron stars at more than $5\sigma$ confidence level. 

It is however important to acknowledge that in the assumption of an accretion disk extending to the ISCO, relativistic reflection probes the size of the ISCO in gravitational radii ($r_g$), which is in turn used to infer the spin of the black hole through the Kerr metric (\citealt{PhysRevLett.11.237}). The Kerr metric describes the spacetime around an uncharged, rotating black hole in the assumption of a vacuum medium, therefore requiring the presence of an event horizon, which is uncharacteristic of neutron stars. Assuming the Kerr metric, the measured spin value of this source of $a=0.86\pm0.02$ implies that the ISCO size that the relativistic reflection models require to explain the observed spectra is $R_{\rm ISCO}\sim2.57\;r_g$. Figure 2 in \cite{2018ApJ...861..141L} illustrates the size of the ISCO of neutron stars of different masses, using different neutron star equations of state. A similar treatment of the ISCO size of neutron stars is performed by \cite{2011MNRAS.415.3247B}. Under the most fortuitous assumption, the minimum size of the ISCO around a neutron star is $\sim4.5\;r_g$, much larger and inconsistent with the measurements of our study, suggesting that the compact object in this system must be a black hole.

Furthermore, a simple estimation of the compactness required solidifies the conclusion. Given the spin measurement, assuming a black hole in the Kerr metric, the compactness at the ISCO is $C=M/R\simeq0.26\;M_\odot/\rm km$, independent of the mass assumption. In comparison, under the simplistic assumption that the system contains a neutron star and an accretion disk that extends to its surface, we can compute the compactness as the mass of the NS divided by its radius. \cite{PhysRevLett.115.161101} find that in order to prevent NS equations of state from breaking causality, for $M=1.4\;M_\odot$, a radius greater than 10.7 km is needed, which leads to $C \lesssim 0.13\;M_\odot/\rm km$. For some typical NS examples, \cite{2021ApJ...918L..28M} find that PSR J0740+6620 has a mass of $M=2.1\;M_\odot$ and a radius $R\simeq13.7\;\rm km$, leading to a compactness $C=0.15$, while \cite{2017A&A...601A.108H} measure the compactness of RX J0720.4-3125 to be $C\simeq0.11$. The compactness required by the relativistic reflection spectrum is much higher than that achievable by neutron stars. The high spin measurement excludes the possibility that the compact object in Swift J1728.9-3613 is a neutron star, providing unequivocal evidence that Swift J1728.9-3613 harbors a black hole. The possibility of a neutron star is similarly, even more significantly rejected when testing higher density models.

Balakrishnan et al. 2023 analyzed multiple avenues of distinguishing between the two types of compact objects. Firstly, in addition to the high spin measurement, superimposing the X-ray colors of Swift J1728.9-3613 onto Figure 8 in \cite{done2003} shows that the source displays behaviour consistent with known black holes, and was measured to have color values that are inaccessible to neutron stars. There is also a lack of evidence for the presence of a neutron star. NICER observations were used to create power spectra, which lacked the high frequency signal that is typically seen in neutron star power spectra (\citealt{belloni2012}). No thermonuclear bursts were detected throughout the observations, and given the total exposure time between Swift, Chandra, NICER, and NuSTAR, the probability that Swift J1728.9-3613 displayed a burst that went undetected is $\sim2\%$. Lastly, a pulsation search in the data returned no results. The results of these tests make it unlikely that the X-ray binary contains a neutron star. Furthermore, \cite{2023MNRAS.519..519S} used similar timing and spectral properties of the source to conclude that it harbors a black hole. They detected two Type-B quasi-periodic oscillations (QPOs) at 5.40 and 5.56 Hz, consistent with a black hole in soft-intermediate state.

\subsection{Initial Black Hole Spin}
By replacing the \texttt{diskbb} component of our model with a more physically accurate description of a thin accretion disk around a Kerr black hole (\texttt{kerrbb} - \citealt{2005ApJS..157..335L}), we can make an inference about the properties of the black hole. We fixed the distance to the black hole to 8.4~kpc estimated by Balakrishnan et al. 2023, the mass of the black hole to $10\;M_\odot$, the normalization of the \texttt{kerrbb} component to 1, and linked the spin and inclination of the inner disk in the \texttt{kerrbb} component to those in \texttt{relxill}, starting from the same best-fit parameters in our baseline model. After fitting, all the parameters of the reflection component remained unchanged, including the black hole spin and inclination of the inner disk. Constraining these parameters maintains the same number of degrees of freedom as in our baseline model, with the quality of the fit worsening by $\Delta\chi^2=1$. 

When fixing all the above mentioned parameters, the shape of the disk component is entirely controlled by the mass accretion rate and the spectral hardening factor. To produce the same shape of the disk component as that predicted by \texttt{diskbb}, the \texttt{kerrbb} component requires a mass accretion rate of $\dot{M}\sim9\times10^{17}\;\rm g/s=1.4\times10^{-8}\;\rm M_\odot/\rm yr$ and a high spectral hardening factor $f_{\rm col}=3.5$. This high value of $f_{\rm col}$ is in conflict with the expected value for accretion disks around stellar-mass black holes in soft states of 1.5-1.9 (\citealt{1995ApJ...445..780S}), but expected for hard states (\citealt{2013MNRAS.431.3510S}). In the assumption of the result of Balakrishnan et al. 2023 that the black hole is associated with the SNR that it is located in, if we assume this estimate for the mass accretion rate to be constant throughout the life of the black hole and by assuming that the age of the black hole is the same as the age of its supernova remnant, on the order of $t=3\times10^4$ years (Balakrishnan et al. 2023), we estimate that the total mass accreted by the black hole in Swift J1728.9-3613 is on the order of $4\times10^{-4}\;\rm M_{\odot}$. 

\citealt{2008ApJ...682..474B} give the equations to calculate the evolution of the spin of a BH of an initial mass and spin, given that it accretes some known amount of mass. These equations were also presented in an appendix in \cite{2000NewA....5..191B} and are derived from the equations of motion on a circular equatorial orbit around a Kerr BH presented first by \cite{1972ApJ...178..347B} (equations 2.12 \& 2.13). As the amount of energy and angular momentum inherited by accreting a unit of mass (and also how much the BH mass and angular momentum increases by adding that energy and angular momentum) changes depending on the BH mass and spin, computing the final BH mass and spin upon accreting some amount of mass is an iterative process. Therefore, writing the equations of the inverse process becomes even more complicated. Thus, we chose to numerically invert the equations and compute the initial spin, given the final BH spin and mass and the amount of mass accreted. 

For a black hole with a final mass of $10\;\rm M_\odot$ and a final spin of $a=0.86$ that throughout its lifetime accreted $4\times10^{-4}\;\rm M_{\odot}$, we found an initial spin smaller by $\Delta a=3.2\times10^{-5}$. The choice of black hole mass in this case has minimal effect, and testing black holes of mass $7\;\rm M_\odot$ or $15\;\rm M_\odot$ changed this final result by at most a factor of 2. An alternative to estimating the amount of mass accreted throughout the lifetime of the BH is to assume that it has been uniformly accreting and reprocessing energy to radiation with some efficiency (in this calculation we assumed $\eta=0.1$) which, observed from the distance of 8.4~kpc that we measure, produces the flux that we observed during the outburst. By extrapolating the best-fit model to the NuSTAR data over the 0.1--300~keV range and by accounting for obscuration along the line of sight, we find that for a source located at a distance of 8.4~kpc, the corresponding luminosity is $3.5\times10^{38}\;\rm erg/s$. At the assumed accretion efficiency, this luminosity corresponds to an accretion rate of $\dot{M}=6.1\times10^{-8}\;\rm M_{\odot}/yr$, which over the entire age of 30,000 years would mean that $1.8\times10^{-3}\;\rm M_{\odot}$ were accreted. Using the same algorithm, we find that this corresponds to an initial spin smaller by $\Delta a=1.4\times10^{-4}$. 

In an even more extreme case, in order to obtain a lower limit on the initial spin of the black hole, we could assume that throughout its lifetime, the black hole accreted matter at the Eddington rate. In this case, for a much more conservative calculation, we assume a black hole with an exaggerated age of an order of magnitude larger than in the previous calculation ($t=10^6$ years). Assuming an efficiency of converting accreted matter to radiation of $\eta=10\%$, we obtain that the black hole would have accreted 2.3\% of its final mass and that in order to have a final spin equal to that we measure, the initial spin would have been smaller by at most $\Delta a=0.02$. By virtue of the assumption of Eddington rate accretion, this estimate is independent of the assumed BH mass. Even under the most conservative choice of parameters for this estimation, the natal spin of the black hole in Swift J1728.9-3613 needs to have been $a_{\rm initial}\geq0.84\pm0.02$.

\section{Discussion} \label{sec:disc}

We analyzed the NuSTAR observation of X-ray binary Swift J1728.9-3613 taken during its 2019 outburst. Modeling the spectra as an absorbed disk blackbody and accounting for relativistic reflection through the \texttt{relxill} model, we were able to measure the spin of the compact object in this X-ray binary, obtaining $a=0.86\pm0.02$. The measured inclination of the inner accretion disk is small, $\rm \theta =<10$ degrees, the iron abundance is subsolar $A_{\rm Fe}=<0.84$, for an accretion disk with moderate ionization $\log(\xi)=3.6\pm0.1$. The high measured spin excludes the possibility for the compact object in the system to be a neutron star, requiring it therefore to be a black hole, in agreement with the conclusions of Balakrishnan et al. 2023. 

Assuming the conclusion of Balakrishnan et al. 2023 that the black hole in this system is likely associated with the supernova remnant G351.9-0.9, we can make estimates for its age by linking it to the age of the supernova remnant. Based on this, we demonstrate that the high value of the spin of the black hole cannot be explained through accretion from a stellar companion and it needs to have been natal.  We find that the black hole must have been born with a spin larger than $0.84\pm0.02$. Thus, whether black holes form via direct collapse (leaving no supernova remnant, such as Cygnus X-1 - \citealt{2003Sci...300.1119M}) or after a neutron star is overwhelmed by gravity, stellar core collapse events can leave black holes with high spin parameters (the spin of the black hole in Cygnus X-1 is $a\geq0.97$ - \citealt{2015ApJ...808....9P}).  

A black hole with an initial spin of $a=0$ needs to accrete 80$\%$ of its mass to increase its spin to $a=0.9$ and an additional $\sim33\%$ of its initial mass to increase the spin from $a=0.9$ to $a=0.98$ (\citealt{1970Natur.226...64B, 2008ApJ...682..474B}). \cite{2015ApJ...800...17F} argue based on a sample of 16 galactic LMXBs that the observed black hole spin can be explained through accretion from the stellar companion and that the entire sample of black holes is consistent with negligible natal black hole spin. Although it is possible that some stellar-mass black holes reach a high spin value through sustained super-Eddington accretion over timescales of $10^{8}-10^{9}$ years (\citealt{2015ApJ...800...17F}), high natal spins indicate that this is not necessary, and that pre-explosion environmental factors such as binarity may drive the configuration of the progenitor star and the character of the black hole. Therefore, while accretion certainly can influence black hole spins, it is possible that the present-day distribution of black hole spins observed in X-ray binaries may provide hints of the relative rates at which core collapse events produce black holes with different spin values in this population. While at this point, no strong conclusion can be drawn regarding the connection between the observed spin distribution in XB and the natal spins of BHs in these systems, future observational and theoretical studies will be able to definitively link the relationship between the two distributions, highlighting the effect of prolonged accretion.

There are a few means by which our results can be tested in the near future.  Balakrishnan et al. 2023 report the detection of the companion star in Swift J1728.9-3613 in infrared bands.  If radial velocity curves can be harnessed to measure the mass of the black hole, a spin measurement using disk continuum models could be obtained using archival X-ray observations. In the near future, pairing the NuSTAR capabilities with IXPE (\citealt{2016SPIE.9905E..17W}) polarization measurements of the reflected radiation from accretion disks will allow placing even tighter constraints on the black hole spins (see, e.g., \citealt{2009ApJ...691..847L, 2011MNRAS.414.2618M}) and the inner disk inclination (see, e.g., \citealt{2022arXiv220609972K}). Higher resolution spectroscopy observations with XRISM (\citealt{2018SPIE10699E..22T})  will allow simultaneous measurements of relativistic reflection and disk winds. These measurements will serve as a stepping stone for future analysis using ATHENA (\citealt{2018SPIE10699E..1GB}), and they may ultimately deliver the most precise and robust measurements.

In the third edition of the Gravitational-Wave Transient Catalog (GWTC-3 - \citealt{2021arXiv211103606T}) there are a total of 90 probable candidates of gravitational wave (GW) events detected from compact object mergers.  Measuring the spins of the black holes in systems measured through GW signals is still difficult, owing to degeneracies in the models used between the mass ratio of the black holes and the effective spin parameter. This continues to be the case despite suggestions that there is no apparent variation of the spin distribution with increasing black hole masses (\citealt{2021arXiv211103634T}). Additionally, works such as \cite{2017PhRvL.119y1103V} have explained that the posterior distribution of spin measurements from GW signals is dependent on the assumed prior distribution. The number of GW signals detected from binary black hole (BBH) mergers is expected to grow substantially in the near future.  It is important to have informed priors on the distribution of black hole spins prior to mergers, and despite the different evolutionary paths, redshifts, metallicities, and mass distributions, the black holes in X-ray binaries are the clearest comparison sample.

Based on GWTC-3, \cite{2021arXiv211103634T} have determined that when treating the BHs in BBH systems as the "more and less rapidly spinning components" instead of the more and less massive components of BBH, the more rapidly spinning components have a spin distribution that peaks at $a\sim0.4$, with 1st and 99th percentiles at $0.07^{+0.05}_{-0.03}$ and $0.8^{+0.08}_{-0.08}$, while the less rapidly spinning components have a distribution centered below $a\leq0.2$ and with 99\% of values below $0.54^{+0.09}_{-0.08}$. Interestingly, \cite{2021arXiv210804821Q} report that the most massive of the black holes in the GW190403\_051519 event has a dimensionless spin of $0.92^{+0.07}_{-0.22}$. While the formation mechanism of BBH is still a topic of debate, the current view is that the entire observed distribution cannot be obtained through a single formation channel (see e.g., \citealt{2021ApJ...910..152Z}). However, the BH population observed through GW signals has preferentially low spins, while the BH spins measured in XB tend to be much larger, inconsistent with the GW population (\citealt{2021arXiv211102935F,2022arXiv221002479D}). It is important to acknowledge that in order to connect the BH spin distributions, the most appropriate sample for comparing the BBH spins are the BHs in HMXB, but due to observational effects, the existing HMXB sample might be biased towards high spins (\citealt{2021PASA...38...56H}).

Recent simulations of BH formation suggests that isolated BHs that form without a supernova remnant most likely have low natal spins due to efficient angular momentum transport (see e.g., \citealt{2019ApJ...881L...1F, 2022MNRAS.511..176A}). However, BH formation in a binary system drastically changes the range of possible natal BH spins, depending both on the properties of the progenitor star and of the binary. Such properties include the efficiency of angular momentum transfer, the chemical structure and rotation rate of the progenitor, the coupling between the core and the envelope of the companion after main sequence evolution, the nature of the supernova event (direct collapse to a BH, a successful supernova explosion, or a failed supernova), interactions between the components of the binary both before the BH formation event (either through mass transfer during a common envelope phase or through tidal interactions between the core of the BH progenitor and its companion) and after the SN event (by influencing the amount of mass and angular momentum accreted by the newly formed BH from an accretion disk) (see e.g., \citealt{2017ApJ...846L..15B, 2018ApJ...862L...3S, 2018A&A...616A..28Q, 2019arXiv190404835B, 2019ApJ...870L..18Q}). This emphasizes the need to better understand the role of binarity in simulations of massive stars and supernova events. Similarly, it is now important for collapsar models to quantify how often black holes are produced in a manner that also leaves a supernova remnant, whether through a neutron star phase with an outward pressure bounce or some action of hyper-Eddington accretion and jet production.

A progenitor that is spun up through tidal interactions in a tight binary can give rise to a BH with high spin (\citealt{2022arXiv220108407F}). Therefore, one possible simplistic approach to explain the observed difference in the two spin distributions is that the two BHs in the BBH merger events observed through GW could be formed early in the binary evolution, when the progenitors are still in wide orbits and tidal interactions do not significantly alter the natal BH spin, leading to the formation of slowly rotating BHs. However, the size of the orbit after BH formation must still allow the newly formed BHs to coalesce within a reasonable timescale. On the other hand, the BHs observed in XB might be formed at later stages of binary evolution, when the two companions are in close orbits. This would cause the progenitor star to experience significant tidal interactions with its companion during the late stages of its evolution and lead to the formation of a rapidly spinning BH. However, the nature of the supernova event must not disrupt the binary, in order for the system to evolve into the observed X-ray binary systems. If true, this would imply that BH formation during different stages of binary evolution, when the progenitors are in wide or tight orbits, can lead to the two seemingly incompatible spin distributions in the two populations of BH. Fully describing the two distributions of BH spins is essential to providing a unified understanding of stellar-mass black hole formation and evolution.

\acknowledgements
We would like to thank the anonymous reviewers, whose comments and suggestions have helped to clarify the content of this paper. We would like to thank Daniel Proga for his suggestions during the early stages of this project. We thank the NuSTAR director, Fiona Harrison, and the mission scheduling team for making this observation. This research has made use of data and software provided by the High-Energy Astrophysics Science Archive Research Center (HEASARC), which is a service of the Astrophysics Science Division at NASA/GSFC, and of the NuSTAR Data Analysis Software (NuSTARDAS) jointly developed by the ASI Science Data Center (ASDC, Italy) and the California Institute of Technology (Caltech, USA).

\newpage

\bibliography{paper}{}
\bibliographystyle{aasjournal}

\appendix

\section{\texttt{relxill} Flavors}\label{sec:relxill}
Different variants of \texttt{relxill} enable one to probe the effects of different processes and geometries. The main two classes within the \texttt{relxill} family are divided according to the assumptions made about the nature of the hard X-ray  corona. Some models assume a ``lamp post" geometry (indicated by the presence of the text ``\texttt{lp}" in the model name), with the parameter $h$ describing the height of the corona above the accretion disk. The other models make no prior assumptions about the geometry of the corona, and model the radiation incident on the accretion disk as a broken power law with radius: $J\propto r^{-q_1} \text{ for } r<r_{\rm break}$, and $J\propto r^{-q_2} \text{ for } r>r_{\rm break}$. This variation of the models can be identified by the lack of ``\texttt{lp}" in the model name.

Within the two coronal geometry classes, different models enable one to probe various different physical mechanisms. For example, \texttt{relxill} and \texttt{relxilllp} model the illuminating flux as a power law of spectral index $\Gamma$ with a high energy cutoff $E_{\rm cut}$ while \texttt{relxillCp} and \texttt{relxilllpCp} model the illuminating flux as an \texttt{nthcomp} Comptonization continuum. While most models assume a density of the accretion disk of $n=10^{15}\;\rm cm^{-3}$, \texttt{relxillD} and \texttt{relxilllpD} enable one to probe disk densities between $n=10^{15}-10^{19}\;\rm cm^{-3}$. Lastly, while most models assume a constant disk ionization $\xi=L/n r^2$ \footnote{$L$ represents the ionizing luminosity, $n$ the number density of the reflector, and $r$ the distance between source and reflector}, the models \texttt{relxilllpion} and \texttt{relxilllpionCp} include a ionization gradient in the accretion disk.

All \texttt{relxill} models include the spin of the compact object $a = cJ/GM^{2}$ where $-0.998 \leq a \leq 0.998$ (\citealt{1974ApJ...191..507T}), the inclination of the inner accretion disk $\theta$, the inner and outer disk radius $r_{\rm in}$ and $r_{\rm out}$, the iron abundance $A_{\rm Fe}$ measured in terms of Solar abundance, and a normalization parameter $\rm norm_{\rm rel}$. Lastly, the models have a reflection fraction parameter $\rm R$, defined as the ratio of the intensity incident on the disk to the intensity escaping to infinity, in the frame of the primary source. If set to negative values, the model only includes reflection features. In our analysis, we used positive values for the reflection fraction, which make the model include the underlying coronal emission in addition to the reflection spectrum. Additionally, we fixed the outer radius of the disk to a value of $r_{\rm out}=990\;r_{\rm g}$ \footnote{$r_{\rm g}=GM/c^2$}, just below the maximum value allowed in the model ($1000\;r_{\rm g}$) and the inner disk radius $r_{\rm in}=r_{\rm ISCO}$, as explained in Section \ref{sec:intro}. 

\section{Inclination experiments}\label{sec:incl}

\begin{figure}[ht!]
    \centering
    \includegraphics[width= 0.65\textwidth]{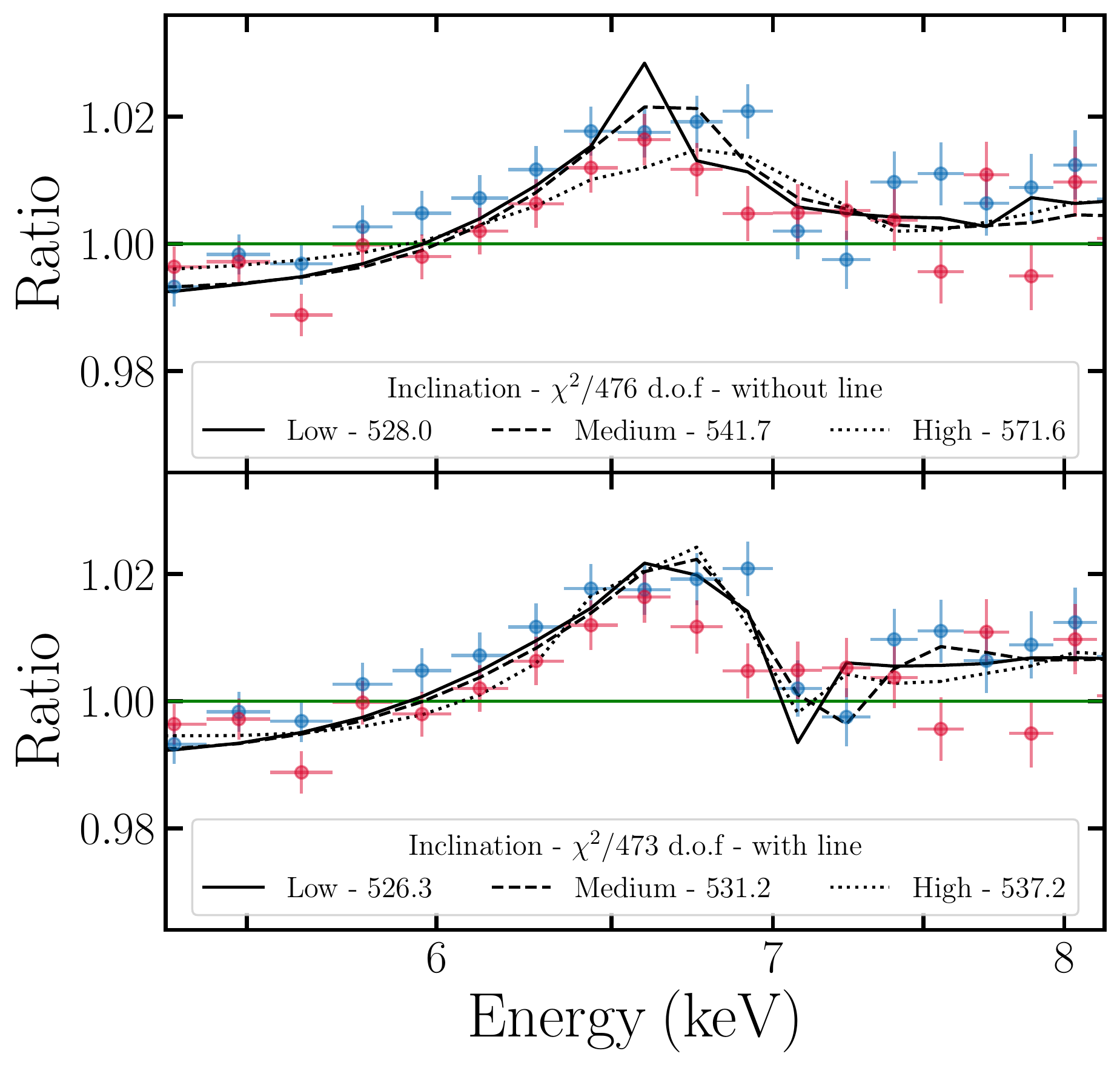}
    \caption{Ratio of the FPMA (blue) and FPMB (red) spectra to the power law model, shown by the green solid line, in the Fe K line region. The solid black line represents the best-fit model when the inclination is constrained to be $\theta\leq30^\circ$, the dashed line represents the model when $30^\circ<\theta<60^\circ$, and the dotted line represents the model with $60^\circ\leq\theta$. The top panel shows the models without the addition of a negative \texttt{gauss} component representing an absorption line around 7.1~keV, while the bottom panel shows the models with a \texttt{gauss} component included.}
    \label{fig:line_shape}
\end{figure}

We tested our model by fitting it to the data with the inclination free, but constrained within three regimes: $\theta\leq30^\circ$, $30^\circ<\rm \theta<60^\circ$, and $60^\circ \leq \rm \theta$. The best fit models for all three inclination regimes are shown in the top panel of Figure \ref{fig:line_shape} as a ratio to the power law model, compared to the data. The quality of the fits increases with a decreasing value of the inclination, as shown in Figure \ref{fig:incl_steppar} as well. 

Perhaps more interesting is the apparent over-prediction of the model when compared to the data around 7.1~keV. To test whether this feature is statistically significant, we introduced a narrow, negative Gaussian component (\texttt{gauss}) to the model representing an absorption feature and re-fitted the data with the inclination constrained to the same three regimes. The best fit models are shown in the bottom panel of Figure \ref{fig:line_shape}, analogous to the models not containing the absorption feature in the top panel. The addition of the \texttt{gauss} component to the models improves the quality of the fits in all three inclination regimes, but does not change the conclusion that a lower inclination is favored. 

\begin{figure}[ht!]
    \centering
    \includegraphics[width= 0.80\textwidth]{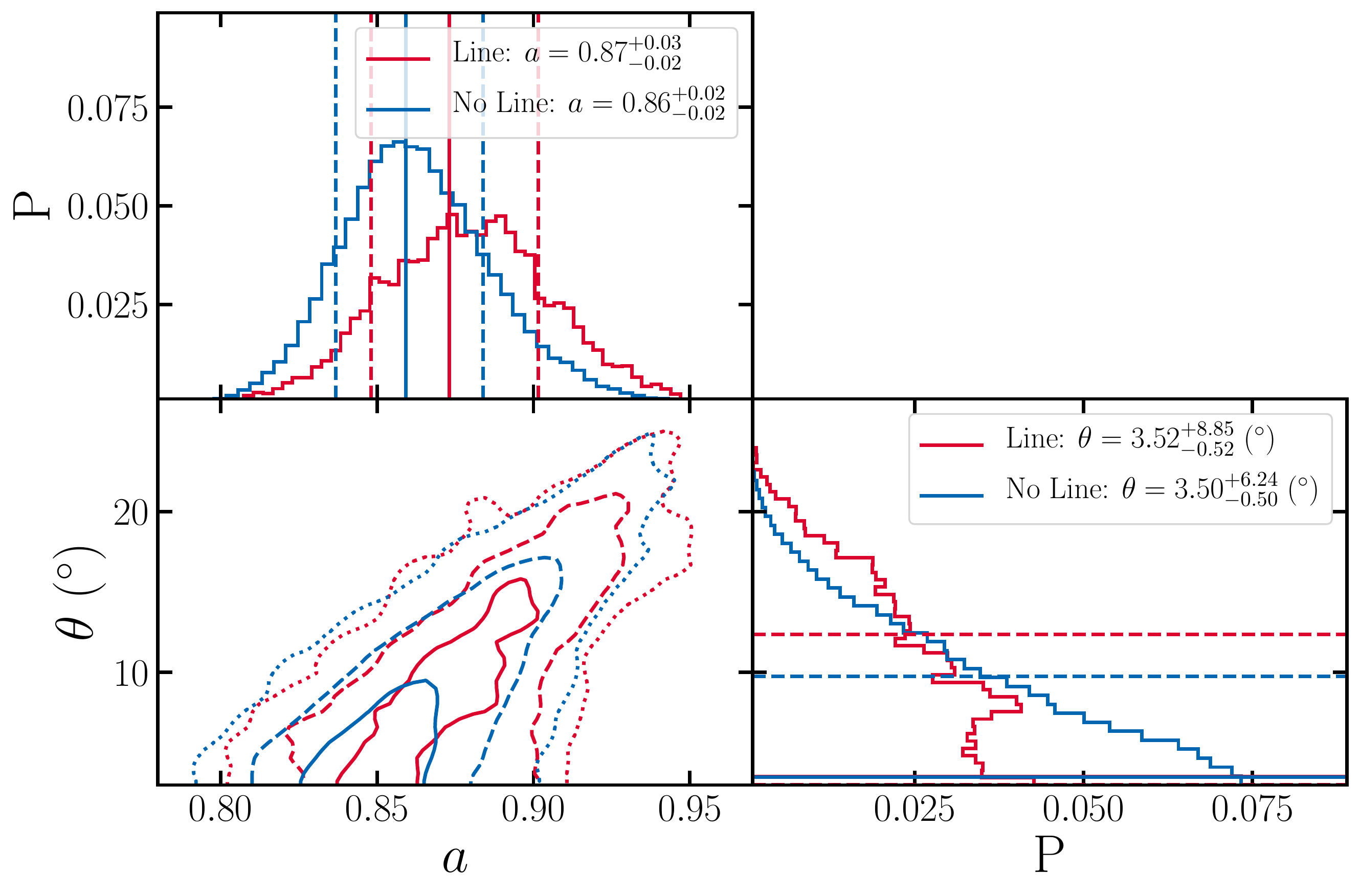}
    \caption{Bottom left panel: 2D histogram of the $a$-$\theta$ parameter space based on the posterior samples in the MCMC analysis for the best fit model with an absorption line at $\sim7$~keV (red) and without one (blue). The solid, dashed, and dotted contours represent the $1\sigma$, $2\sigma$, and $3\sigma$ confidence intervals respectively. The top left and bottom right panels show the 1D histograms of the posterior distribution in the MCMC analysis for spin and inclination with (red) and without (blue) including an absorption line in the model. The vertical solid lines represent the mode of the distributions and the vertical dashed lines represent the $\pm1\sigma$ credible regions.}
    \label{fig:spin_incl_line_noline}
\end{figure}

The addition of the negative \texttt{gauss} component at 7.1~keV improves the quality of the fit by $\Delta \chi^2=1.7$ at the cost of three extra parameters. Given this value, intuitively one would expect that the addition of the component is not statistically significant. However, to test this, we computed the Akaike Information Criterion (AIC) and the Bayesian Information Criterion (BIC) for the model before and after the addition of the \texttt{gauss} component as follows:
\begin{equation}
    AIC=Nln(\chi^2/N)+2N_{var}
\end{equation}
\begin{equation}
    BIC=Nln(\chi^2/N)+ln(N)N_{var} \\ 
\end{equation}

where $N$ is the number of data bins and $N_{var}$ is the number of variables in the model. Without the addition of the line, $AIC=65.67$ and $BIC=128.62$. After including the \texttt{gauss} component, $AIC=70.09$ and $BIC=145.62$. The increase in both AIC and BIC suggests that the addition of the absorption feature is indeed not statistically favored. For completeness, we computed the F-test probability for the addition of the line, despite it not being an accurate way to quantify the significance of a line (\citealt{2002ApJ...571..545P}) and obtained a p-value of 0.68, again suggesting that the absorption line is not statistically significant. While similar absorption features are not uncommon among similar objects (see e.g. \citealt{2020ApJ...900...78D}), the low inclination disfavors absorption along the line of sight, as disk winds are expected to be equatorial (\citealt{2012MNRAS.422L..11P}). Another possible, though unlikely, source of absorption would be parts of the SNR along the line of sight, but improved spectroscopy of the SNR will help determine whether velocities required to produce an absorption line at this energy are present in the SNR. While the preferred inclination remains low both with and without an absorption line around 7~keV, more importantly, the predicted spin remains roughly unchanged. We ran an MCMC analysis similar to that described in section \ref{sec:MCMC} for the best performing model that includes a free absorption \texttt{gauss} component around 7~keV. Figure \ref{fig:spin_incl_line_noline} 1D and 2D contours similar to those shown in Figure \ref{fig:spin_incl_chi} for the two models, highlighting that the measurements are mostly unaffected by the addition of an absorption feature.

\begin{figure*}[ht!]
    \centering
    \includegraphics[width= 0.95\textwidth]{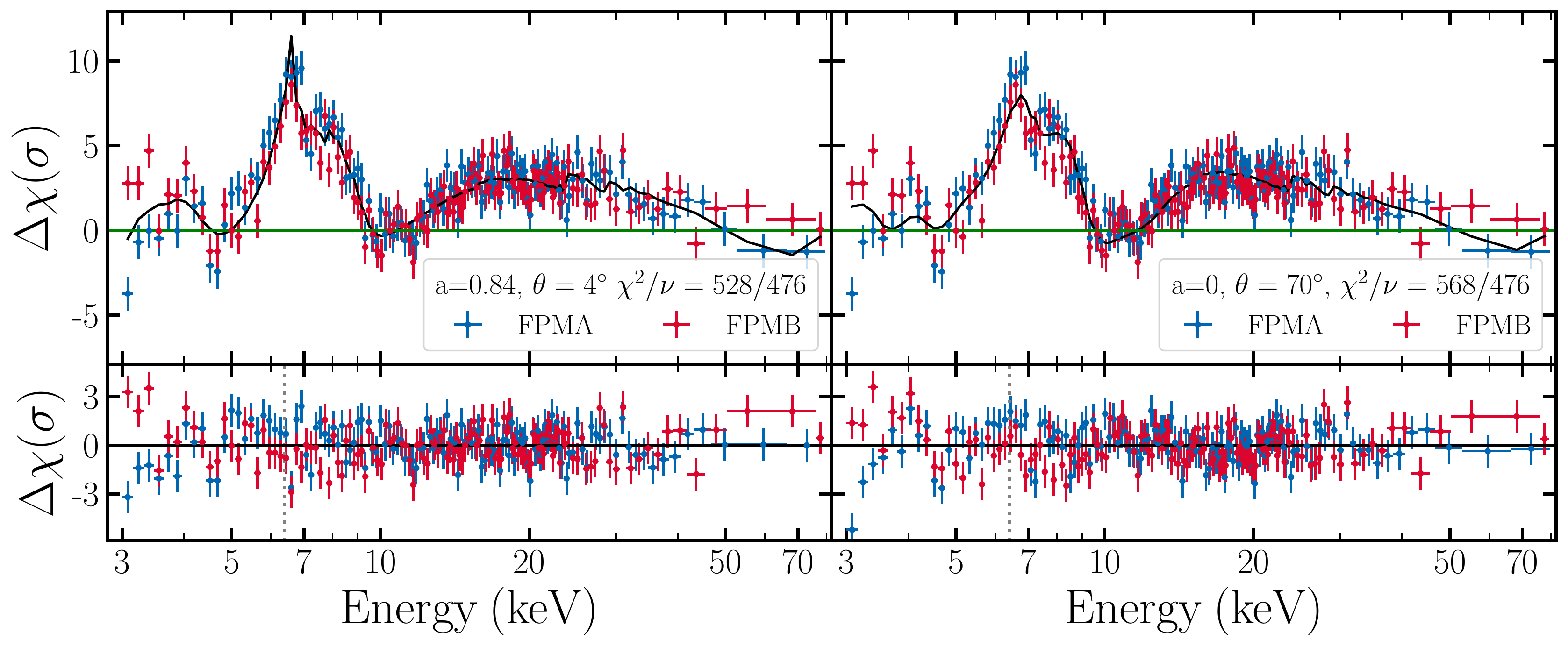}
    \caption{Top: residuals in terms of $\sigma$ produced when fitting the NuSTAR FPMA (blue) and FPMB (red) spectra with \texttt{const*TBabs*(diskbb+powerlaw)}, while ignoring the data between 5--9~keV and 12--40~keV. The green line represents the model. Fitting the residuals with a model accounting for reflection (black line) with low inclination and high spin (left) and high inclination and low spin (right). The bottom panels show the residuals of the spectra when fit with the two reflection models presented in the top panels. The vertical, gray dotted line shows the position of the neutral Fe K$\alpha$ line at 6.4~keV. The low spin, high inclination fit fails to properly model the shape of the broadened Fe K line.}
    \label{fig:incl_residuals}
\end{figure*}

Since the MCMC analysis described in section \ref{sec:MCMC} does not cover the entire parameter space explored by e.g., the analysis leading to the results shown in Figure \ref{fig:incl_steppar}, we decided to further quantify the significance of our results. To test the robustness of the low inclination measurement, we compared the fit that produces a low inclination to the fit that produces the best statistic with a high inclination ($\theta \sim 75^\circ$ - see Figure \ref{fig:incl_steppar}). In the top panels of Figure \ref{fig:incl_residuals} we show the residuals in terms of $\sigma$ produced when fitting the spectra with \texttt{const*TBabs*(diskbb+powerlaw)}, while ignoring the data between 5--9~keV and 12--40~keV, in order to highlight the reflection features. The solid black lines in the top panels in Figure \ref{fig:incl_residuals} represent the best models when recovering a low inclination (left) and a high inclination (right), while the blue and red points represent the FPMA and FPMB spectra respectively. It is important to note that instead of being the same solution with different inclination values, the two models shown represent entirely different solutions, measuring not only different inclinations, but different spins too. The low inclination, high spin measurement (left) produces a fit better by $\Delta \chi^2=40$ for no change in the number of degrees of freedom when compared to the high inclination, low spin fit (right). Since the two fits have the same number of free parameters, the fit producing a worse statistic can be interpreted as a local minimum, and when shifting parameters from that solution and refitting, the fit converges to the low inclination, high spin measurement rather than back to the previous local minimum. 

The bottom panels in Figure \ref{fig:incl_residuals} show the residuals of the models highlighted in the top panels, with the vertical gray dashed lines representing the position of the neutral Fe K$\alpha$ line at 6.4~keV. The main differences between the two solutions are most apparent in the Fe K band and below 5~keV. The high inclination fit clearly fails to capture the complexity of the spectra in the Fe K band. To test the effects of the difference between the two spectra below 5~keV on the fit and the spin and inclination measurements, we reran the fits by first completely ignoring the data below 5~keV of both spectra. In this case, the high inclination, low spin solution returns $\chi^2/\nu=439.41/450$, being worse by $\Delta \chi^2=11$ when compared to the low inclination, high spin solution which returns $\chi^2/\nu=428.45/450$, making the low inclination result still statistically significant. Since information is being lost when ignoring energies below 5~keV, we refit the spectra in the entire 3--79~keV band, but instead of allowing for a multiplicative constant between the two detectors, we allowed each spectra to have independent normalizations of the \texttt{diskbb} and \texttt{relxill} components, while linking all other parameters. This produces $\chi^2/\nu=520.71/475$ and $\chi^2/\nu=537.58/475$ for the two solutions, with the low inclination, high spin solution still being preferred with $\Delta \chi^2=17$ for no change in the number of degrees of freedom. The observed equivalent width of the Fe K line and the reflection fraction decrease with increasing inclination due to extra absorption and scattering of photons reflected off the disk at high angles (\citealt{2000PASP..112.1145F}). Due to these effects, observing systems with low inclinations is not surprising, but rather expected.

The low inclination suggests that if a jet were still currently present, it would be pointed directly towards us. The presence of a jet is backed by the detection of a radio signal from the source using MeerKAT, 4 days before the NuSTAR observation (\citealt{2019ATel12522....1B}). While it would be difficult to infer the impact or absence of a jet in this configuration, in the case in which the corona can be thought of as the base of a jet, this would imply an observation directly through the corona.  It was shown that this configuration would have the effect of blurring the reflection, weakening it, and potentially even leading to lower measured abundances (\citealt{2017ApJ...836..119S}). We attempted to model this effect by modifying our best fit model to include the convolution component \texttt{simplcut} (\citealt{2017ApJ...836..119S}). When fixing the scattering fraction in the \texttt{simplcut} component to 1 in order to simulate a high amount of scattering, the fits become worse by $\Delta \chi^2 \sim 200$ due to an inability of fitting the shape of the Fe K line, and the spin is unconstrained. When allowing the scattering fraction to vary, the fit prefers scattering fractions $\approx 3\times10^{-3}$ and the fit converges to the same solution as before the addition of the extra component to the model, with a minimal decrease in $\chi^2$ statistic ($\Delta\chi^2\sim0.3$).  

We measured the inclination of the inner regions of the accretion disk with respect to the line of sight. As this is a young system, it is possible that the equatorial plane of the black hole and the inner accretion disk have not yet aligned with the outer regions of the accretion disk and therefore with the orbital inclination (see e.g., \citealt{2008MNRAS.387..188M}), suggesting that any inclination measurements derived from orbital motion of the stellar companion do not necessarily reflect the orientation of the inner disk. As natal kicks cannot significantly alter the magnitude and direction of the spin of a new-born black hole (\citealt{2020MNRAS.495.2179S}), the natal spin and orientation of the rotation axis of a black hole has to be set by the angular momentum inherited from the progenitor star during the final stages of the collapse. However, recent studies argue for the presence of inclined spin axes of BHs at their birth, both in X-ray binaries (e.g., \citealt{2022Sci...375..874P}) and in BBH mergers (e.g., \citealt{2022arXiv220502541T}). Further observations of both the inner regions of the disk and of the supernova remnant that Swift J1728.9-3613 is located in would be required in order to infer the presence or effects of a jet as a method of probing the orientation of the rotation axis of the black hole and therefore the inclination of the inner regions of the accretion disk.

\section{Corner plot}\label{sec:corner}

\begin{figure}[ht!]
    \centering
    \includegraphics[width=0.95 \textwidth]{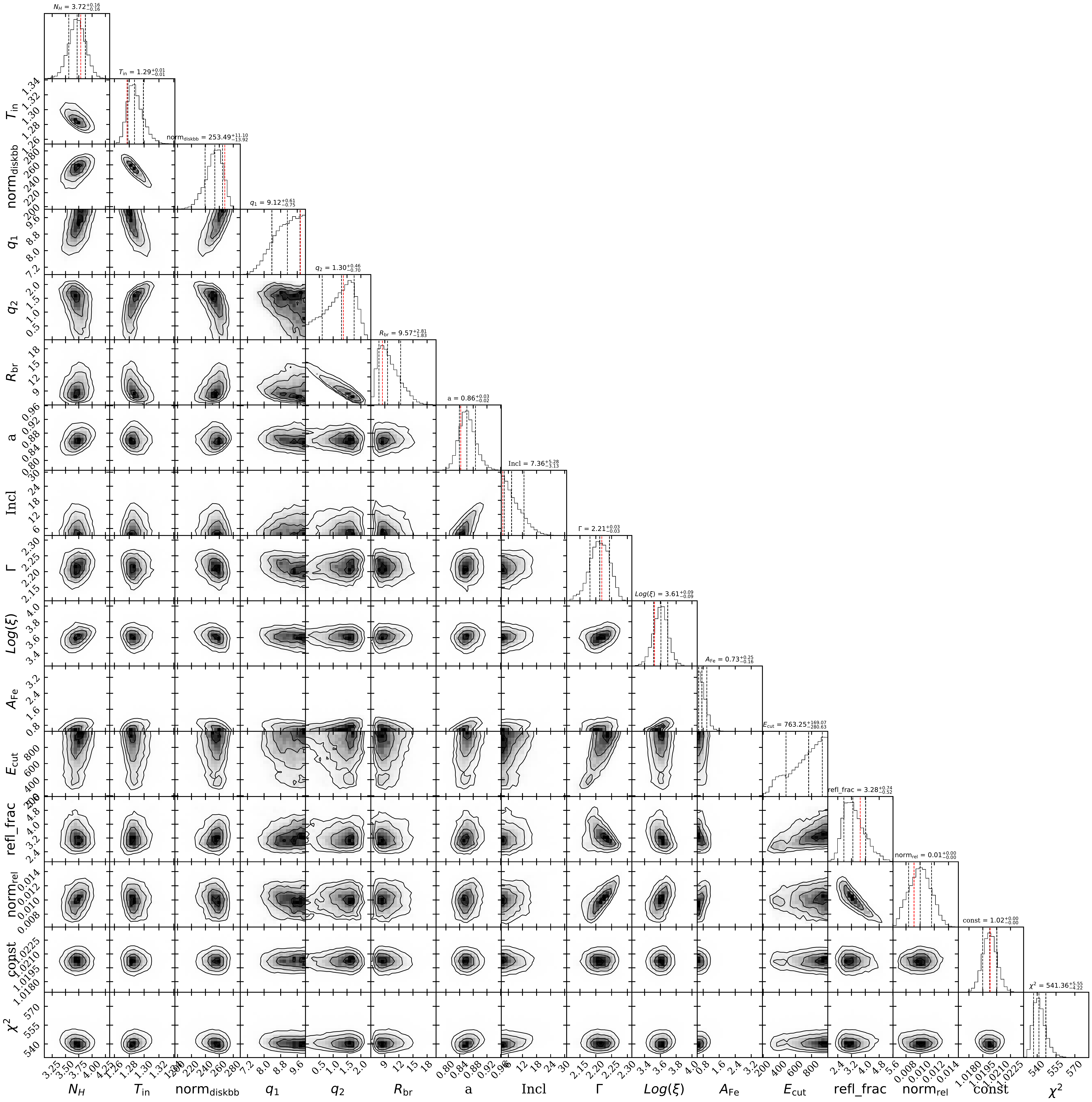}
    \caption{The complete corner plot of the MCMC analysis. The vertical red lines in the 1D posterior distributions represent the values around which Gaussian proposal distributions were generated and used to initialize the walkers in the MCMC run. }
    \label{fig:corner_full}
\end{figure}

Figure \ref{fig:corner_full} shows the complete corner plot generated from the posterior samples resulting from the MCMC analysis. The numbers reported above the 1D distributions represent the median of the posterior sample and the $\pm 1 \sigma$ confidence intervals on the median. We note that these numbers could be different from the values that we report throughout the paper - the mode of the posterior distributions and the $1\sigma$ credible region. For a discussion regarding this choice, see Section 2.2 in \cite{2022arXiv221002479D}. The contours in the plot represent the $1\sigma$, $2\sigma$, and $3\sigma$ confidence intervals in the 2D posterior distribution for each parameter combination. The vertical red lines in the 1D posterior distributions represent the values around which Gaussian proposal distributions were generated and used to initialize the walkers in the MCMC run. In the case of $A_{\rm Fe}$ and $E_{\rm cut}$, the red line is not visible in the plot as it coincides with one of the hard limits for the parameters, $A_{\rm Fe}=0.5$ and $E_{\rm cut}=1000\;\rm keV$. The priors for the parameters are uniform in the parameter range allowed by the model components. 

Based on Figure \ref{fig:corner_full}, there is a negative correlation between the temperature of the \texttt{diskbb} component and its normalization, indicating that this simple, Newtonian interpretation of an accretion disk does not capture the extent of relativistic processes present in an accretion disk around a compact object. As an alternative to the \texttt{diskbb}, we replaced it with \texttt{ezdiskbb} (\citealt{2005ApJ...618..832Z}), which assumes zero torque at the inner disk radius, and \texttt{kerrbb} (\citealt{2005ApJS..157..335L}) which describes the emission from a thin accretion disk while accounting for general relativistic effects. Using \texttt{ezdiskbb} produces a fit worse by $\Delta \chi^2=5$. When using \texttt{kerrbb} and linking the spin and inclination of the \texttt{kerrbb} component to that of \texttt{relxill}, the fit becomes slightly better by $\Delta \chi^2=0.5$, but the model also introduces more complexity through a larger number of free parameters. Most importantly, in both cases, the best-fit reflection component remains roughly unchanged, measuring similar inclinations and spins. Nevertheless, as the parameters of the \texttt{diskbb} component and the spin of the compact object do not show any degeneracy in the MCMC analysis and as this model produces a good fit, the \texttt{diskbb} component suffices in isolating relativistic reflection from direct emission from the disk. 

Another immediately obvious negative correlation between parameters is present between the breaking radius $R_{\rm br}$ of the broken power law emissivity of the corona and the outer emissivity index $q_2$. Once again, this is not unexpected, based on the definition of the two quantities and this correlation appears to have no effect on the other parameters. The possible degeneracy between the inclination and the spin is explored in section \ref{sec:MCMC}. Lastly, the reflection fraction parameter $\rm R$ and the component normalization $\rm norm_{\rm rel}$ appear to be negatively correlated. This has previously been observed in similar measurements (see e.g., \citealt{2020ApJ...900...78D}), but as this does not affect the spin measurement, it was not further explored.

\end{document}